\documentclass[aps,titlepage,12pt,superscriptaddress]{revtex4}
\usepackage{epsfig}
\usepackage{float}
\usepackage{graphicx}
\usepackage{times}
\usepackage{color}
\begin{document}

\title{Competitions between prosocial exclusions and punishments in finite populations}

\author{Linjie Liu}

\author{Xiaojie Chen}
\email{xiaojiechen@uestc.edu.cn} \affiliation{School of Mathematical
Sciences, University of Electronic Science and Technology of China,
Chengdu 611731, China
}

\author{Attila Szolnoki}
\email{szolnoki@mfa.kfki.hu} \affiliation{Institute of Technical
Physics and Materials Science, Centre for Energy Research, Hungarian
Academy of Sciences, P.O. Box 49, H-1525 Budapest, Hungary}

\begin{abstract}\noindent
\\
Prosocial punishment has been proved to be a powerful mean to
promote cooperation. Recent studies have found that social
exclusion, which indeed can be regarded as a kind of punishment, can
also support cooperation. However, if prosocial punishment and
exclusion are both present, it is still unclear which strategy is
more advantageous to curb free-riders. Here we first study the
direct competition between different types of punishment and
exclusion. We find that pool (peer) exclusion can always outperform
pool (peer) punishment both in the optional and in the compulsory
public goods game, no matter whether second-order sanctioning is
considered or not. Furthermore, peer exclusion does better than pool
exclusion both in the optional and in the compulsory game, but the
situation is reversed in the presence of second-order exclusion.
Finally, we extend the competition among all possible sanctioning
strategies and find that peer exclusion can outperform all other
strategies in the absence of second-order exclusion and punishment,
while pool exclusion prevails when second-order sanctioning is
possible. Our results demonstrate that exclusion is a more powerful
strategy than punishment for the resolution of social dilemmas.
\end{abstract}

\maketitle

\noindent

\section*{Introduction}
\noindent Cooperation is widespread in our world, which has a
fundamental role on the evolution of human civilization
\cite{axelod_r_m06,taylor_m87,searle_j10,boyd_r09,chase_i_d80}.
However, cooperation is vulnerable to be invaded by selfish
individuals who are always maximizing their short-term and immediate
interests. Thus how to overcome such individuals is a vital task for
the emergence of cooperation in a population
\cite{hardin_g_s68,komorita_s_s09,miller_d_t99}. Several mechanisms,
such as spatial reciprocity, reputation, wisdom of groups, and
costly punishment, have been demonstrated to be effective for
cooperators to fight against defectors
\cite{clutton_b_t02,fu_f08,chen_x15,szolnoki_a12,boyd_r10,santos_prl05,perc_sr15}.
Staying at the last option, costly punishment has received
considerable attention in the last decade because of its importance
and widespread prevalence in human societies
\cite{henrich_j06,chen_xj15,helbing_d10,chen_x_j14}. By using public
goods game (PGG), which is a standard metaphor of social dilemmas,
many theoretical and experimental studies have shown that prosocial
punishment can reduce the number of free-riders and encourage the
majority of individuals to contribute to the common pool
\cite{denant-Boemont_l07,sigmund_k10,shinada_m07,traulsen_a12}.

As an alternative incentive tool to prevent free-riders exploiting
community effort, social exclusion can also be observed in human
societies \cite{gruter_m86,feinberg_m14,molden_d_c09}. It is based
on the idea that convicted offenders are denied certain rights and
benefits of citizenship or membership of joint ventures
\cite{travis_02}. Accordingly, individuals who are identified to
violate the rule or jeopardize others' common interests could be
excluded from the community
\cite{unit_s_e01,twenge_j_m01,byrne_d05,masclet_d03}. In this way
exclusion serves as a sort of institution to tame defectors not to
exploit others. Previous studies have shown that social exclusion
can increase social sensitivity
\cite{twenge_j_m07,dewall_c_n11,tuscherer_t15,eisenberger_n_i06,bernstein_m_j12}
and induce a positive impact on cooperation when partners are fixed
\cite{maierrigaud_f_p10,ouwerkerk_j_w05}. Recently, Sasaki and
Uchida introduced peer exclusion into the PGG and established a
game-theoretical model to study the evolution of social exclusion by
using replicator equations in infinite populations
\cite{sasaki_t13}. They found that peer exclusion can overcome two
shortages of peer punishment: first, a rare punishing cooperator
barely subverts the asocial society of free-riders; second, natural
selection often eliminates punishing cooperators in the presence of
non-punishing cooperators (namely, second-order free-riders).
Subsequently, Li et al. \cite{li_k15} studied the comparison between
peer exclusion and pool exclusion, and claimed that peer excluders
can overcome pool excluders if the exclusion costs are small and
excluders can dominate the whole population in a suitable parameters
range in the presence of second-order free-riders. Note that pool
excluder, similarly to pool punisher, pays a fixed, permanent cost
before contributing to the public goods to maintain an
institutionalized mechanism for punishing exploiters.

To summarize our present knowledge, both prosocial exclusion and
prosocial punishment have been proved to be effective ways for
promoting cooperation, but their systematic comparison is still
missing. Indeed, the mentioned works ~\cite{sasaki_t13,li_k15} have
compared their independent impacts on the cooperation level, but the
consequence of their simultaneous presence is still unexplored. It
remained unclear which strategy is more evolutionary advantageous if
both exclusion and punishment are simultaneously available for
individuals in the population. We wonder if exclusion or punishment
is a better way to curb free-riding. How does their relation change
if second-order sanctioning is also possible? In the latter case
non-punishing individuals or those who deny contribution to the cost
of exclusion may also be punished. Furthermore, we also wonder
whether peer punishment (peer exclusion) or pool punishment (pool
exclusion) is more efficient individual strategy to control
transgressors for a higher well-being.

Motivated by these open problems, in this study we focus on the
competition between prosocial exclusion and punishment in finite
populations who play the PGG. We first investigate the direct
competition between pool exclusion and pool punishment, and
demonstrate that pool exclusion has the evolutionary advantage over
pool punishment both in the optional and in the compulsory PGG, no
matter whether second-order exclusion and punishment are considered
or not. We then investigate the competition between peer exclusion
and peer punishment, and find that peer exclusion is evolutionarily
advantageous over peer punishment both in the optional and in the
compulsory PGG, independently of the choice of second-order
sanctioning. Third, we study the competition between pool exclusion
and peer exclusion, and observe that peer exclusion can outperform
pool exclusion both in the optional and in the compulsory PGG if
second-order exclusion is ignored, while the situation is reversed
in the presence of second-order exclusion. Finally, we investigate
the full competition of all previously mentioned strategies, such as
pool exclusion, peer exclusion, pool punishment, and peer
punishment. As our main observation, it turns out that peer
exclusion is the most advantageous strategy in the absence of
second-order exclusion and punishment, but pool exclusion
outperforms other strategies when second-order sanctioning is
possible.

\section*{Model}

We consider the standard PGG in a finite, well-mixed population with
size $M$. In each round of the game, $N\geq2$ individuals are
selected randomly from the population to form a group for
participating in a one-shot game. Then, each individual in the group
decides whether or not to contribute an amount of cost $c$ to the
common pool. The individual who is willing to contribute is called a
cooperator, and the individual who does not contribute is called a
defector. In the optional PGG we also consider a third option, a
strategy which gives up participating in the game, hence is called
as a loner. The latter strategy has a constant payoff $\sigma$ which
is not affected by others. The sum of the contributions to the
common pool is multiplied by the enhancement factor $r$ ($1<r<N$),
and then equally allocated among all individuals who participated in
the game no matter they contributed or not. In agreement with
previous works ~\cite{hauert_c02,hauert_c07}, if only one individual
participates in the game then her income equals with $\sigma$.

In the second stage of the game exclusion or/and punishment is
considered where both related strategies contribute $c$ to the
common pool. By following Refs. \cite{sigmund_k10,sasaki_t15} peer
punishers impose a fine $\beta$ on each free-rider in their group at
a cost $\gamma$. Accordingly, each defector will be fined an amount
$\beta N_W$, where $N_W$ is the number of peer punishers in the
group. Pool punishers, however, pay a permanent cost $G$ to the
punishment pool beforehand.
If there exist defectors in the group,
they will be fined an amount $BN_V$, where $N_V$ is the number of
pool punishers in the group.
It simply means that the additional cost of pool punisher is independent of the number of defectors in the group, while the related cost of peer punisher is proportional to the presence of defectors.
If considering second-order punishment,
second-order free-riders (individuals who contribute to the game but
do not bear the extra cost of punishment) will be fined the same
amount \cite{sigmund_k10}.

When exclusion is applied we follow conceptually similar protocol as
for punishment. Here exclusion serves as a sort of institution to
prevent defectors to exploit other group members. Hence the role of
excluder can be viewed as a sentinel who alarms other group members
about the danger of defectors. Evidently, such an extra effort
requires additional cost which is paid by excluder player.
Consequently, a peer excluder does not only contribute $c$ to the
public goods game but also pay a cost $c_{E}$ after every defector
in the group to prevent them collecting benefit from the public
goods sharing. In stark contrast to peer exclusion, pool excluders
pay a permanent cost $\delta$ to maintain the institution of
exclusion which will block defectors to gain benefit from PGG in the
presence of pool excluders. As previously, in case of the
second-order exclusion, second-order free-riders (individuals who do
not take the extra cost of exclusion) will also be excluded.

In order to study the evolutionary dynamics, we use the so-called pairwise comparison rule with the mutation-selection process \cite{szabo_pr07,traulsen_a07}. According to this protocol at each time step a randomly chosen player $i$ may change her strategy. We consider the possibility of mutation, hence the player adopts a randomly chosen available strategy with probability $\mu$. Alternatively, which happens with probability $1-\mu$, a player $i$ tries to imitate a randomly chosen player $j$ with a probability
\begin{eqnarray}
f(\Pi_{j}-\Pi_{i})=\frac{1}{1+\exp^{-\kappa(\Pi_{j}-\Pi_{i})}} \,\,.
\end{eqnarray}
Here $\Pi_{i}$ and $\Pi_{j}$ are the collected payoffs of the mentioned players $i$ and $j$, while $\kappa$ characterizes the intensity of selection. In the $\kappa \rightarrow \infty$ strong imitation limit the more successful player $j$ always succeeds in enforcing her strategy to player $i$, but never otherwise. On the other hand, $\kappa \rightarrow 0$ indicates the so-called weak selection limit where strategy adoption becomes random independently of the payoff values. In between these extremes, at a finite value of $\kappa$, it is likely that a better performing player $j$ is imitated, but it is still not impossible to adopt her strategy when performing worse.

In the following we consider four different scenarios when punishment and exclusion compete and we compute the resulting stationary distribution of all available strategies. We suppose a well-mixed finite population where all players interact with each other randomly.
To make comparison with previous works easier we have adopted notations for variables by earlier works \cite{sigmund_k10, sasaki_t13}.
Accordingly, let $X$ denote the number of cooperators who contribute to the public pool, but do not bear the cost of punishment or exclusion; $Y$ the number of defectors who contribute neither to PGG nor to the sanctions; $Z$ the number of loners; $V$ the number of pool punishers; $W$ the number of peer punishers; $F$ the number of pool excluders; and $E$ the number of peer excluders. The whole size of population is denoted by $M$ and $N$ randomly chosen individuals are offered to form a group and establish a joint enterprise.
In the next section we present the results of the more complex optional PGG while further details and results for the simplified compulsory PGG game are summarized in the Supplementary Information (SI).

\section*{Results}
\noindent {\subsection*{Competition between pool exclusion and pool punishment}}

We first study the direct competition between pool exclusion and
pool punishment in the optional PGG. In this scenario, there are
five available strategies in the population fulfilling the
constraint $X+Y+Z+F+V=M$. We assume that $0<\sigma<rc-c-\delta$ and
$0<\sigma<rc-c-G$, which ensure that a punisher or excluder can get
higher profit than a loner if there is more than one participant in
the group. In the absence of second-order exclusion and punishment
only defectors are sanctioned by punishment or/and exclusion. In
Fig.~\ref{fig1}(a) we plot the long-run frequencies for each
strategy which determine the stationary distribution of all
available strategies in dependence of imitation strength $\kappa$.
We find that for $\kappa<10^{-4}$ the frequencies of the five
competing strategies are identical due to the practically random
imitation process. As we increase the strength of imitation then all
the five strategies can survive and coexist. More precisely, in the
perfect imitation limit the system evolves towards a homogeneous
state, where the flips between almost homogeneous states are
triggered by rare mutations. (A representative trajectory of
evolution can be seen in Fig.~1 in Ref.~\cite{sigmund_k10}.) In this
way the presented frequencies of stationary states are calculated
from the time average of frequencies for competing strategies. As
Fig.~\ref{fig1}(a) suggests pure cooperators form the highest
portion who can enjoy the benefit of exclusion and punishment
without paying their costs. Interestingly, the second largest
population is formed by pool-excluders followed by defectors and
loners, while pool-punishers can make up the smallest fraction, or
can be detected with the smallest probability. This result suggests
that pool exclusion is more effective against defection and has an
evolutionary advantage over pool-punishment strategy. In particular,
in the strong imitation ($\kappa \rightarrow \infty$) limit the
long-run frequencies in the $[X, Y, Z, F, V]$ subpopulations are
$\frac{3}{8}, \frac{3}{16}, \frac{1}{8}, \frac{4M+5}{16M+32}$, and
$\frac{M+5}{16M+32}$, respectively (for further details, see
Section~$1$ in SI).

The comparison of stationary strategies in Fig.~\ref{fig1}(a)
emphasizes that it is better to cooperate but also to avoid the
additional cost of sanctions. Needless to say, if everyone chose
this option then we would face the original dilemma. To minimize
this undesired consequence of ``second-order free-riders" we may
penalize those who refuse participating in the sanctioning process.
In particular, when second-order exclusion and punishment are
considered, we assume that pool-excluders drive out all strategies
who do not contribute to the exclusion pool. In parallel,
pool-punishers will also punish those who do not bear the additional
cost of punishment pool. By applying this scenario, we observe that
the long-run frequencies in the $[X, Y, Z, F, V]$ subpopulations are
$[0, 0, 0, 1, 0]$ for $\kappa>10^{-3}$, as shown in
Fig.~\ref{fig1}(b). In other words, when second-order sanctioning is
allowed pool-excluders prevail and all other strategies extinct
during the evolutionary process. To answer our original question
both panels plotted in Fig.~\ref{fig1} highlight that pool-exclusion
is a more advantageous strategy than pool-punishment independently
of second-order sanctioning is considered or not. We should stress
that our observation remains valid for a broad range of parameter
interval including the level of the punishment fine or the group
size. Furthermore, conceptually identical conclusion can be made if
the compulsory PGG is assumed where loner strategy cannot compete
(the related results are presented in Sec.~$2$ of SI).

{\bfseries \subsection*{Competition between peer exclusion and peer punishment}}

In this subsection, we investigate the competition between peer
punishment and peer exclusion in the optional PGG. According to this
scenario there are also five available strategies in the population
whose fractions fulfill $X+Y+Z+E+W=M$. In the absence of
second-order punishment and exclusion, we assume that
$0<c_{E}<\frac{rc-c}{N-1}$, which ensures that a single peer
excluder can invade a group of all defectors. In Fig.~\ref{fig2}(a)
we present the long-run frequencies of the five strategies as a
function of the imitation strength $\kappa$. As expected, for weak
selection, when $\kappa<10^{-4}$, the frequencies of the five
strategies are practically identical because of the random strategy
updating. By increasing the imitation strength $\kappa$ peer
excluder strategy becomes gradually dominant and occupies the
majority of the population. Peer punisher strategy can only reach
the second best position in the rank of strategies. In the strong
imitation limit the long-run frequencies for the $[X, Y, Z, E, W]$
subpopulations are $[\frac{6}{5M+23}, \frac{3}{5M+23},
\frac{2}{5M+23}, \frac{3M+6}{5M+23}, \frac{2M+6}{5M+23}]$ (see
Section.~$3$ of SI). This suggests that the number of peer excluders
is about $1.5$ times larger in time average than the second best
peer punishers for large population size, hence demonstrating the
superiority of the former strategy.

If second-order sanctioning is allowed then the relation of
sanctioning strategies becomes even more unambiguous. Interestingly,
in this case excluders do not only ostracize pure cooperators but
also punishers who refuse to contribute to the cost of exclusion.
But the penalty works also in the reversed direction because
punishers lower the payoff of both cooperation and exclusion
strategies. The consequence of this mutual sanctioning is summarized
in Fig.~\ref{fig2}(b), which suggests that peer exclusion prevails
and gives no space for any other strategies. This observation
supports our previous conclusion about the effectiveness of
exclusion that is not restricted to pool strategies, but is still
valid for peer strategies. We stress that this conclusion remains
unchanged if we release the restriction for the value of $c_E$,
which means that for $c_E>\frac{rc-c}{N-1}$ peer exclusion has still
evolutionary advantage over peer punishment. The border within this
observation is valid can be extended further because peer exclusion
outperforms peer punishment in the compulsory PGG, no matter whether
second-order sanctioning is applied or not (for more details see
Section~$4$ in SI).

{\bfseries \subsection*{Competition between pool exclusion and peer
exclusion}}

In the following subsection we compare the peer- and pool-exclusion
strategies in the optional PGG, which are proved to be more
effective than their punishing mates in the previously studied
cases. Here, there are five available strategies in the population
whose fractions fulfill the constraint $X+Y+Z+F+E=M$. For their
proper comparison we assume that their costs remain below the
previously established limit, that is $0<c_{E}<\frac{rc-c}{N-1}$ and
$ 0<\sigma<rc-c-\delta$. First, we consider the case when
second-order exclusion is not allowed, hence both excluder
strategies penalize pure defectors only. Fig.~\ref{fig3}(a)
illustrates that peer excluder strategy becomes dominant as we
gradually increase the imitation strength. All the other strategies
can share a reasonable portion only at an intermediate value
$\kappa$. If the imitation strength exceeds the threshold $\kappa
>10^{-1}$ then the long-run frequencies of defectors, loners, and
pool excluders are close to zero, and only cooperators can coexist
with peer excluders. In particular, the fractions of $[X, Y, Z, F,
E]$ strategies in the strong imitation ($\kappa\rightarrow \infty$)
limit are $[\frac{9}{6M+25}, \frac{3}{6M+25}, \frac{2}{6M+25},
\frac{2}{6M+25}, \frac{6M+9}{6M+25}]$ (further details can be seen
in Section~$5$ of SI). These results suggest that peer excluder
strategy is able to dominate the whole population in the absence of
second-order exclusion.

Interestingly, the outcome of evolutionary trajectory is completely reversed if
second-order exclusion is considered. In this case, in strong
agreement with a previous work where peer- and pool-punisher
strategies were compared \cite{sigmund_k10}, pool excluders are
capable to crowd out peer excluders. The result of this competition
is summarized in Fig.~\ref{fig3}(b) where the long-run frequencies
for each strategy are plotted. In the strong imitation limit the
victory of pool excluders is total, yielding  $[0, 0, 0, 1, 0]$
fractions for $X,Y,Z,F$, and $E$ strategies respectively. As in the
previous cases, these results remain valid if the compulsory PGG is
played. Here, in the absence of loners, peer-excluders dominate when
second-order exclusion is not considered yielding $[\frac{4}{3M+11},
\frac{2}{3M+11}, \frac{1}{3M+11}, \frac{3M+4}{3M+11}]$ values for
the competing $X, Y, F$, and $E$ strategies in the strong imitation
limit (details can be found in Sec.~6 of SI). When second-order
exclusion is possible then the pool excluder strategy prevails in
close agreement with the result of optional PGG.

{\bfseries \subsection*{Competition between prosocial exclusions and
punishments}}

The pair comparison of competing strategies may provide a first
guide about their relations, but the presence of a third party could
be a decisive factor, which may completely rearrange the ranks of
competitors. To clarify this possibility in the following we explore
the simultaneous competitions of all previously studied strategies.
Namely, we consider an optional PGG where seven strategies, namely
pure cooperator, defector, loner, peer excluder, peer punisher, pool
excluder, and pool punisher are present. As in the previous cases,
we first consider the option when second-order sanctions are not
applied hence only defectors suffer from the presence of excluders
and punishers. Fig.~\ref{fig4}(a) summarizes our results, which
suggest that ``peer-sactioning" strategies are the most effective,
but more importantly, peer excluders can dominate the population. In
this way the dominance of peer excluders over peer punishers is not
disturbed by the presence of other sanctioning strategies such as
pool excluders or pool punishers. As Fig.~\ref{fig4}(a) shows all
the other strategies become irrelevant in the strong imitation
limit. In particular, the long-run frequencies of $X,Y,Z,E,W,F$ and
$V$ subpopulations are $\frac{45}{163+35M}$, $\frac{15}{163+35M}$,
$\frac{6}{163+35M}$, $\frac{20M+45}{163+35M}$,
$\frac{15M+45}{163+35M}$, $\frac{18M+7}{(163+35M)(3M+2)}$, and
$\frac{3M+7}{(163+35M)(3M+2)}$, respectively (further details can be
seen in Section~$7$ of SI). This result suggests that only the
sanctioning strategies survive in the large population limit where the majority of individuals are peer excluders in most of the time.

In the next logical step we consider the case when second-order
sanctioning is possible. This option offers an extremely complex
food-web between competing strategies, because practically all
sanctioning strategies try to beat all the others. For instance,
pool excluders ostracize not only defectors and simple cooperators,
but also peer excluders, peer punishers, and pool punishers. In this
``almost everybody beats everybody else" battle the final victor is
pool excluder strategy. This case is plotted in Fig.~\ref{fig4}(b)
where the resulting fractions of the strategies are $[0, 0, 0, 0, 0,
1, 0]$ in the strong imitation limit (further details can be seen in
Sec.~7 of SI).

To close this section we briefly summarize the results of the
compulsory PGG where 6 competing strategies remain. The details of
the calculation can be found in Sec.~8 of SI. In the absence of
second-order exclusion and punishment, we find that the behaviour is
conceptually similar to the one we observed for the optional PGG.
Here peer excluders and peer punishers perform the best, but all the
other strategies survive at intermediate strength of imitation. In
the strong imitation limit the resulting fractions of $X,Y,E,W,F$,
and $V$ strategies are $[\frac{6}{5M+22}, \frac{3}{5M+22},
\frac{3M+6}{5M+22}, \frac{2M+6}{5M+22}, \frac{3M+1}{(5M+22)(3M+2)},
\frac{1}{(5M+22)(3M+2)}]$, which suggests that only sanctioning
strategies survive in the large-population limit. When second-order
exclusion and punishment are possible then we get back the result
obtained previously for the optional PGG: only pool excluders
survive for strong enough imitation strength.

\section*{Discussion}

\noindent Penalizing free-riders whose behaviour threaten the
collective efforts seems almost inevitable. But which sanctioning
tool shall we apply to reach our goal efficiently? To punish them by
lowering their payoffs or to deny their rights to enjoy the benefit
of public goods? The answer could be even more complicated because
both peer and pool sanctioning can be used. While peer punishers and
peer excluders invest an extra cost only in the presence of
defectors, pool punisher and pool excluder strategies apply a
permanent effort to maintain the sanctioning institutions. Based on
previous works both punishment and exclusion seem to be appropriate
methods \cite{sigmund_k10,abdallah_s14}, but their systematic
comparison has not been done yet.

In this work, we have thus studied the competitions between costly
punishments and exclusions in finite populations playing the PGG by
using different scenarios. For a fair comparison we have applied
equally high cost of punishment and exclusion. We have found that
peer exclusion is always favored by natural selection when it
competes with peer punishment both in the optional and in the
compulsory PGG, independently of second-order punishment and
exclusion are considered or not. Conceptually similar findings have
been obtained for pool exclusion when it directly competes with pool
punishment. Furthermore, when peer exclusion competes with pool
exclusion, peer exclusion wins in the absence of second-order
exclusion, while pool exclusion prevails when second-order exclusion
is applied. Lastly, we have also explored the most complex option
when all four sanctioning methods compete with the pure cooperator,
defector, and loner strategies. In the latter case peer exclusion is
proved to be the most viable tool in the absence of second-order
punishment and exclusion, while pool exclusion prevails when
second-order sanctioning is allowed. To sum up, the systematic
comparison of sanctioning strategies highlights that exclusion is
always a more effective way to control free-riders than punishment,
but the absence or the presence of second-order sanctioning could be
a decisive factor, because the former condition supports peer
exclusion while the latter option helps pool exclusion strategy to
prevail.

We would like to stress that our finding is robust and remains valid
in a broad range of model parameters (some representative plots are
given in Sec. $9$ of SI). For instance, if we increase the
punishment fine by fixing the cost of punishment then the
superiority of exclusion is still not in danger. In general, if the
fine is not unrealistically high and the cost of exclusion does not
exceed the cost of punishment then exclusion strategy always
performs better similarly to the cases we discussed earlier. Indeed,
we have verified that in the absence of second-order sanctioning the
exclusion strategy still has an evolutionary advantage over the
punishment strategy, no matter an enhanced fine value applied by
peer and pool punishers. To give an example, the outcomes remain
conceptually intact when the punishment fine exceeds eight times the
punishment cost. But if second-order sanctioning is applied at such
severe punishment then the advantage of excluders diminishes because
their payoff becomes negative, which implies the victory of
punishers. However, we should note that applying such a severe
punishment is not an attractive feature when humans qualify
potential social partners \cite{rockenbach_pnas11}. We have also
considered different group sizes and found that it has no
significant role in the competition of sanctioning strategies (this
is demonstrated clearly in Fig.~S7 of SI).

In order to provide a convenient framework for studying the
competitions between costly exclusions and punishments, we focused
on the option when free-riders are always exiled in the presence of
excluders who have to bear the related cost. A further step could be
when this sanction is not perfect and exclusion happens in a
probabilistic manner \cite{sasaki_t13,li_k15}. Indeed, previous
works emphasized the value of probabilistic sanctioning
\cite{szolnoki_jtb13,chen_x_j14}, which opens promising avenue for
future studies. Our work can be also extended where the error of
perception, i.e. defectors are identified with some ambiguity, or
the error of punishment or exclusion are also considered. In the
latter cases innocent players are punished or excluded from the
joint venture by mistake. To consider anti-social punishment and
anti-social exclusion may also open interesting research avenue to
explore the effectiveness of exclusion
\cite{hauser_jtb14,dossantos_prsb15,szolnoki_prsb15}. Lastly, we
note that our calculation is restricted to the simplest, well-mixed
population because of the extremely high number of competing
strategies. However, it is a frequently discussed fact that in
structured populations, where interaction topology is considered,
the evolutionary outcomes could be significantly different from
those presented for mean-field systems
\cite{masuda_prsb07,perc_m13,pinheiro_njp12,szolnoki_pre11,szolnoki_prx13,chen_pre15}.
Therefore, we expect similar exciting new observations from related
efforts which will hopefully make our understanding more accurate.

\clearpage

\noindent \textbf{Acknowledgments} \\
This research was supported by the National Natural Science
Foundation of China (Grants No. 61503062) and the Fundamental
Research Funds of the Central Universities of China.

\noindent \\ \textbf{Author contributions} \\
The authors designed and performed the research as well as wrote the
paper.

\noindent \\ \textbf{Competing financial interests} \\
The authors declare no competing financial interests.

\clearpage

\begin{figure}
\centering
\caption{{\bf The competition between pool exclusion and pool
punishment in the optional PGG.} Without second-order exclusion or
punishment, shown in panel~(a), the frequency of pool excluders is
significantly higher than the portion of pool punishers, but all
strategies can coexist in time average. In the presence of
second-order exclusion and punishment, presented in panel~(b), pool
excluders prevail and dominate in the strong selection limit.
Parameters: $N=5$, $r=3$, $c=1$, $\mu=10^{-6}$, $\sigma=1$, $M=100$,
$\delta=0.4$, and $G=B=0.4$. }\label{fig1}
\end{figure}

\begin{figure}
%\centering
%\includegraphics[width=1.0\textwidth]{fig2.eps}
\caption{{\bf The competition between peer exclusion and peer
punishment in the optional PGG.} In the absence of second-order
sanctioning, shown in panel~(a), both strategies survive but peer
exclusion dominates. If second-order exclusion and punishment are
applied then peer excluders prevail, as shown in panel~(b). Both the
cost and fine of punishment are equal. Parameters: $N=5$, $r=3$,
$c=1$, $\mu=10^{-6}$, $\sigma=1$, $M=100$, $c_{E}=0.4$, and
$\beta=\gamma=0.4$}\label{fig2}
\end{figure}

\begin{figure}
%\centering
%\includegraphics[width=1.0\textwidth]{fig3.eps}
\caption{{\bf The competition between pool exclusion and peer
exclusion in the optional PGG.} In the absence of second-order
exclusion, peer excluders prevail, shown in panel~(a). Panel~(b)
shows the opposite case which happens if second-order exclusion is
applied. Parameters: $N=5$, $r=3$, $c=1$, $\mu=10^{-6}$, $\sigma=1$,
$M=100$, and $c_{E}=\delta=0.4$.}\label{fig3}
\end{figure}

\begin{figure}
%\centering
%\includegraphics[width=1.0\textwidth]{fig4.eps}
\caption{{\bf The competition between different types of social
exclusion and costly punishment in the optional PGG.} When
strategies can penalize defectors only, shown in panel~(a), then all
strategies coexist in time average for weak strength of imitation,
but in most of the time peer excluders form the majority of the
population for other imitation strength values. Panel~(b) shows the
case when second-order sanctioning is possible. Here pool excluders
prevail and conquer the whole population. Parameters: $N=5$, $r=3$,
$c=1$, $\mu=10^{-6}$, $\sigma=1$, $M=100$, $c_{E}=\delta=0.4$,
$\beta=\gamma=0.4$, and $B=G=0.4$.}\label{fig4}
\end{figure}

\end{document}

\title{Supplementary Information: Competitions between prosocial exclusions and punishments in finite populations}

\author{Linjie Liu}
\affiliation{School of Mathematical Sciences, University of
Electronic Science and Technology of China, Chengdu 611731, China}

\author{Xiaojie Chen}
\affiliation{School of Mathematical Sciences, University of
Electronic Science and Technology of China, Chengdu 611731, China}

\author{Attila Szolnoki}
\affiliation{Institute of Technical Physics and Materials Science,
Centre for Energy Research, Hungarian Academy of Sciences, P.O. Box
49, H-1525 Budapest, Hungary}

\maketitle

In the Supplementary Information (SI), we provide the specific
analysis for the strategy distribution in the finite population with
size $M$ by elaborating eight different scenarios. In order to
calculate the stationary distribution of more than two strategies in
our study, we consider that the mutation probability $\mu$ is
sufficiently small that is in agreement with previous works
\cite{abdallah_s14,hauert_c07,sigmund_k07,traulsen_a06}.
Consequently, the population will never contain more than two
different strategies simultaneously, and the population can evolve
into a homogeneous state where all individuals adopt the same
strategy because the time between two mutation events is long
enough. This assumption allows us to approximate the evolutionary
dynamics by means of an embedded Markov chain whose states
correspond to the different homogeneous states of the population. If
the number of available strategies is denoted by $d$  then the state
transition matrix which describes the different transition
probabilities for the population to move from one state to the other
is given by
\begin{eqnarray*}
\textbf{A}=[a_{hq}]_{d\times d},
\end{eqnarray*}
where $a_{hq}$ is the probability that the system switches from the
$h$ state (where all individuals adopt strategy $h$) to the $q$
state (where all individuals adopt strategy $q$) after the emergence
of a single mutation). Here $a_{hq}=\frac{\rho_{hq}}{d-1}$ if $h\neq
q$ and $a_{hq}=1-\sum_{q\neq h}\frac{\rho_{hq}}{d-1}$ otherwise,
where $\rho_{hq}$ is the fixation probability that a single
individual with strategy $q$ takes over a resident population of
individuals with strategy $h$. The stationary distribution of the
population can be calculated from the average fraction of time that
population spends in each of the homogeneous states. Technically, it
is given by the normalized left eigenvector of the eigenvalue $1$ of
the transition matrix $\textbf{A}$.

The fixation probability $\rho_{hq}$ can be calculated as follows.
We assume that in a finite population with size $M$ there are $H$
individuals using strategy $h$ and $Q=M-H$ individuals using
strategy $q$. Then the probability that the number of individuals
who use strategy $h$ increases/decreases by one is
\begin{eqnarray*}
T^{\pm}(H)=\frac{H}{M}\frac{M-H}{M}\frac{1}{1+exp^{\mp
\kappa(\Pi_{hq}-\Pi_{qh})}},
\end{eqnarray*}
where $\Pi_{hq}$ and $\Pi_{qh}$ are the average payoffs of
individuals with strategy $h$ and with strategy $q$, depending on
the numbers $H$ and $Q$. Correspondingly, the fixation probability
$\rho_{hq}$ can be expressed by
\begin{align*}
\rho_{hq}&=\frac{1}{1+\sum_{m=1}^{M-1}\prod_{Q=1}^{m}\frac{T^{-}(H)}{T^{+}(H)}}\nonumber\\
&=\frac{1}{1+\sum_{m=1}^{M-1}exp^{[\kappa\sum_{Q=1}^{m}(\Pi_{hq}-\Pi_{qh})]}}.
\end{align*}

\noindent

\section*{1 Competition between pool exclusion and pool punishment in the optional PGG}
\noindent

First, we calculate the average payoffs for each strategy in a
finite population assuming that $M$ individuals consist of $X$
cooperators, $Y$ defectors, $Z$ loners, $F$ pool excluders, and $V$
pool punishers. Thus the probability to find no pool excluders in
the population is
\begin{eqnarray*}
\frac{\binom{M-F-1}{N-1}}{\binom{M-1}{N-1}}.
\end{eqnarray*}
Accordingly, the average payoff for cooperators is given by
\begin{eqnarray*}
\lefteqn{\Pi_{X}=\frac{\binom{Z}{N-1}}{\binom{M-1}{N-1}}\sigma+[1-\frac{\binom{Z}{N-1}}{\binom{M-1}{N-1}}]\{[1-\frac{\binom{M-F-1}{N-1}}{\binom{M-1}{N-1}}](rc-c)}\\
&&+\frac{\binom{M-F-1}{N-1}}{\binom{M-1}{N-1}}\sum_{k=0}^{N-1}\sum_{j=0}^{N-k-1}\frac{\binom{M-Y-Z-F-1}{N-j-k-1}\binom{Y}{j}\binom{Z}{k}}{\binom{M-F-1}{N-1}}[\frac{r(N-j-k)c}{N-k}-c]\},
\end{eqnarray*}
where $k$ represents the number of loners, and $j$ represents the number of defectors in the group. Defectors are not only excluded by pool excluders, but also punished by pool punishers. Thus, the average payoff for defectors is
\begin{eqnarray*}
\lefteqn{\Pi_{Y}=\frac{\binom{Z}{N-1}}{\binom{M-1}{N-1}}\sigma+[1-\frac{\binom{Z}{N-1}}{\binom{M-1}{N-1}}]\{[1-\frac{\binom{M-F-1}{N-1}}{\binom{M-1}{N-1}}][-\frac{(N-1)VB}{M-1}]+\frac{\binom{M-F-1}{N-1}}{\binom{M-1}{N-1}}}\\
&&\sum_{i=0}^{N-1}\sum_{j=0}^{N-i-1}\sum_{k=0}^{N-i-j-1}\frac{\binom{M-Y-X-Z-F}{N-i-j-k-1}\binom{Y-1}{j}\binom{Z}{k}\binom{X}{i}}{\binom{M-F-1}{N-1}}[\frac{r(N-j-k-1)c}{N-k}-(N-i-j-k-1)B]\},
\end{eqnarray*}
where $i$ represents the number of cooperators. The average payoff for loners is
\begin{eqnarray*}
\Pi_{Z}=\sigma.
\end{eqnarray*}
The average payoff for pool excluders is
\begin{eqnarray*}
\Pi_{F}=\frac{\binom{Z}{N-1}}{\binom{M-1}{N-1}}\sigma+[1-\frac{\binom{Z}{N-1}}{\binom{M-1}{N-1}}](rc-c-\delta).
\end{eqnarray*}
Finally, the pool punishers' payoff is
\begin{eqnarray*}
\Pi_{V}&=&\frac{\binom{Z}{N-1}}{\binom{M-1}{N-1}}\sigma+[1-\frac{\binom{Z}{N-1}}{\binom{M-1}{N-1}}]\{[1-\frac{\binom{M-F-1}{N-1}}{\binom{M-1}{N-1}}](rc-c-G)\\
&&+\frac{\binom{M-F-1}{N-1}}{\binom{M-1}{N-1}}\sum_{j=0}^{N-1}\sum_{k=0}^{N-j-1}\frac{\binom{M-F-Y-Z-1}{N-j-k-1}\binom{Y}{j}\binom{Z}{k}}{\binom{M-F-1}{N-1}}[\frac{r(N-j-k)c}{N-k}-c-G]\}.
\end{eqnarray*}

In what follows, we calculate the elements of the transition matrix
$\textbf{A}$ to determine the stationary distribution of the
population. First, we calculate the payoff expressions $\Pi_{hq}$ of
strategy type $h$ competing against type $q$ for all the possible
pairs, which are the followings.
\begin{eqnarray*}
\Pi_{XY}&=&\sum_{i=0}^{N-1}\frac{\binom{X-1}{i}\binom{M-X}{N-i-1}}{\binom{M-1}{N-1}}[\frac{r(i+1)c}{N}-c]
=\frac{rc}{N}[\frac{(N-1)(X-1)}{M-1}+1]-c,\\
\Pi_{YX}&=&\sum_{i=0}^{N-1}\frac{\binom{X}{i}\binom{M-X-1}{N-i-1}}{\binom{M-1}{N-1}}\frac{ric}{N}
=\frac{rc}{N}\frac{(N-1)X}{M-1},\\
\Pi_{XZ}&=&\frac{\binom{Z}{N-1}}{\binom{M-1}{N-1}}\sigma+[1-\frac{\binom{Z}{N-1}}{\binom{M-1}{N-1}}](rc-c)
=rc-c-\frac{\binom{Z}{N-1}}{\binom{M-1}{N-1}}(rc-c-\sigma),\\
\Pi_{ZX}&=&\Pi_{ZY}=\Pi_{ZF}=\Pi_{ZV}=\sigma,\\
\Pi_{XF}&=&\Pi_{XV}=rc-c,\\
\Pi_{FX}&=&\Pi_{FY}=\Pi_{FV}=rc-c-\delta,\\
\Pi_{VX}&=&\Pi_{VF}=rc-c-G,\\
\Pi_{YF}&=&0,\\
\Pi_{YZ}&=&\frac{\binom{Z}{N-1}}{\binom{M-1}{N-1}}\sigma,\\
\Pi_{YV}&=&\frac{(N-1)V}{M-1}(\frac{rc}{N}-B),\\
\Pi_{FZ}&=&rc-c-\delta-\frac{\binom{Z}{N-1}}{\binom{M-1}{N-1}}(rc-c-\delta-\sigma),\\
\Pi_{VY}&=&\frac{rc}{N}[\frac{(N-1)(V-1)}{M-1}+1]-c-G,\\
\Pi_{VZ}&=&rc-c-G-\frac{\binom{Z}{N-1}}{\binom{M-1}{N-1}}(rc-c-G-\sigma).
\end{eqnarray*}
Based on the above payoff expressions, we can give the embedded
Markov chain describing the transition between cooperators ($X$),
defectors ($Y$), loners ($Z$), pool excluders ($F$), and pool
excluders ($V$) as
\begin{equation*}
\left(
\begin{array}{ccccc}
I_{X} & \frac{\rho_{XY}}{4} & \frac{\rho_{XZ}}{4}  & \frac{\rho_{XF}}{4} & \frac{\rho_{XV}}{4}\\
\frac{\rho_{YX}}{4} & I_{Y} & \frac{\rho_{YZ}}{4} & \frac{\rho_{YF}}{4}  & \frac{\rho_{YV}}{4}\\
\frac{\rho_{ZX}}{4} & \frac{\rho_{ZY}}{4} & I_{Z} & \frac{\rho_{ZF}}{4} & \frac{\rho_{ZV}}{4}\\
\frac{\rho_{FX}}{4} & \frac{\rho_{FY}}{4} & \frac{\rho_{FZ}}{4} & I_{F} & \frac{\rho_{FV}}{4}\\
\frac{\rho_{VX}}{4} & \frac{\rho_{VY}}{4} & \frac{\rho_{VZ}}{4} & \frac{\rho_{VF}}{4} & I_{V}\\
\end{array}
\right),
\end{equation*}
where $I_{K}=1-\sum_{K\neq L} \frac{\rho_{KL}}{4}$, and $K, L\in\{X, Y, Z, F, V\}$. Correspondingly, we can determine the long-run frequency for each strategy as a function of the imitation strength $\kappa$  [see Fig. 1(a)].

In the strong imitation limit the embedded Markov chain describing the transitions matrix is simplified as
\begin{equation*}
\left(
\begin{array}{ccccc}
\frac{3}{4} & \frac{1}{4} & 0 & 0 & 0\\
0 & \frac{1}{2} & \frac{1}{4} & \frac{1}{4} & 0\\
\frac{1}{8} & 0 & \frac{5}{8} & \frac{1}{8} & \frac{1}{8}\\
\frac{1}{4} & 0 & 0 & \frac{3}{4}-\frac{1}{4M} & \frac{1}{4M}\\
\frac{1}{4} & 0 & 0 & \frac{1}{4M} & \frac{3}{4}-\frac{1}{4M}\\
\end{array}
\right).
\end{equation*}
Accordingly, the stationary distribution is $[\frac{3}{8},
\frac{3}{16}, \frac{1}{8}, \frac{4M+5}{16M+32},
\frac{M+5}{16M+32}]$, which demonstrates that all the five
strategists can coexist, but the frequency of pool punishers is
significantly lower than pool excluders.

In the presence of second-order exclusion and punishment, we assume that pool excluders exclude pool punishers, pure cooperators, and defectors. And pool punishers punish pool excluders, pure cooperators, and defectors. The average payoffs for defectors and loners are not changed. But the average payoff for cooperators is given by
\begin{eqnarray*}
\lefteqn{\Pi_{X}=\frac{\binom{Z}{N-1}}{\binom{M-1}{N-1}}\sigma+[1-\frac{\binom{Z}{N-1}}{\binom{M-1}{N-1}}]\{[1-\frac{\binom{M-F-1}{N-1}}{\binom{M-1}{N-1}}][-c-\frac{(N-1)VB}{M-1}]+}\\
&&\frac{\binom{M-F-1}{N-1}}{\binom{M-1}{N-1}}\sum_{i=0}^{N-1}\sum_{k=0}^{N-1-i}\sum_{p=0}^{N-1-i-k}\frac{\binom{M-X-Y-F-Z}{p}\binom{Z}{k}\binom{X-1}{i}\binom{Y}{N-i-k-p-1}}{\binom{M-F-1}{N-1}}[\frac{rc(i+p+1)}{N-k}-c-pB]\},
\end{eqnarray*}
where $p$ represents the number of pool punishers in the group. The average payoff for pool excluders is given by
\begin{eqnarray*}
\Pi_{F}&=&\frac{\binom{Z}{N-1}}{\binom{M-1}{N-1}}\sigma+[1-\frac{\binom{Z}{N-1}}{\binom{M-1}{N-1}}]\sum_{i=0}^{N-1}\sum_{k=0}^{N-1-i}\sum_{p=0}^{N-1-i-k}\frac{\binom{M-X-Y-V-Z-1}{N-i-k-p-1}\binom{Z+Y}{k}\binom{X}{i}\binom{V}{p}}{\binom{M-1}{N-1}}[\frac{rc(N-k)}{N-i-p-k}\\
&&-c-\delta-pB].
\end{eqnarray*}
Last, the payoff for pool punishers is
\begin{eqnarray*}
\lefteqn{\Pi_{V}=\frac{\binom{Z}{N-1}}{\binom{M-1}{N-1}}\sigma+[1-\frac{\binom{Z}{N-1}}{\binom{M-1}{N-1}}]\{[1-\frac{\binom{M-F-1}{N-1}}{\binom{M-1}{N-1}}](-c-G)+}\\
&&\frac{\binom{M-F-1}{N-1}}{\binom{M-1}{N-1}}\sum_{i=0}^{N-1}\sum_{k=0}^{N-1-i}\sum_{p=0}^{N-1-i-k}\frac{\binom{M-X-Y-F-Z-1}{p}\binom{Z}{k}\binom{X}{i}\binom{Y}{N-i-k-p-1}}{\binom{M-F-1}{N-1}}[\frac{rc(i+p+1)}{N-k}-c-G]\}.
\end{eqnarray*}

Again, the transition matrix between cooperators ($X$), defectors
($Y$), loners ($Z$), pool excluders ($F$), and pool punishers ($V$)
can be simplified in the strong imitation limit as
\begin{equation*}
\left(
\begin{array}{ccccc}
\frac{1}{2} & \frac{1}{4} & 0 & \frac{1}{4} & 0\\
0 & \frac{1}{2} & \frac{1}{4} & \frac{1}{4} & 0\\
\frac{1}{8} & 0 & \frac{5}{8} & \frac{1}{8} & \frac{1}{8}\\
0 & 0 & 0 & 1 & 0\\
0 & 0 & 0 & \frac{1}{4} & \frac{3}{4}\\
\end{array}
\right).
\end{equation*}
The resulting stationary distribution is $[0, 0, 0, 1, 0]$, which
suggests that pool excluders prevail [results are summarized in Fig.
1(b)].

\section*{2 Competition between pool exclusion and pool punishment in the compulsory PGG}
\noindent

In this section we assume that the population contains $X$
cooperators, $Y$ defectors, $F$ pool excluders, and $V$ pool
punishers from which $N$ individuals are selected randomly to play
the PGG. In the absence of second-order exclusion and punishment, a
cooperator obtains the average payoff as
\begin{align*}
\Pi_{X}&=[1-\frac{\binom{M-F-1}{N-1}}{\binom{M-1}{N-1}}](rc-c)+\frac{\binom{M-F-1}{N-1}}{\binom{M-1}{N-1}}\sum_{i=0}^{N-1}\frac{\binom{M-Y-F-1}{i}\binom{Y}{N-i-1}}{\binom{M-F-1}{N-1}}[\frac{r(i+1)c}{N}-c]\\
&=rc[1-\frac{\binom{M-F-1}{N-1}}{\binom{M-1}{N-1}}\frac{(N-1)Y}{N(M-F-1)}]-c.
\end{align*}
The payoff for defectors is
\begin{eqnarray*}
\Pi_{Y}&=&[1-\frac{\binom{M-F-1}{N-1}}{\binom{M-1}{N-1}}]\sum_{p=0}^{N-1}\frac{\binom{M-V-1}{N-p-1}\binom{V}{p}}{\binom{M-1}{N-1}}(-pB)+\\
&&\frac{\binom{M-F-1}{N-1}}{\binom{M-1}{N-1}}\sum_{j=0}^{N-1}\sum_{p=0}^{N-1-j}\frac{\binom{M-Y-F-V}{N-j-p-1}\binom{V}{p}\binom{Y-1}{j}}{\binom{M-F-1}{N-1}}[\frac{r(N-j-1)c}{N}-pB]\\
&=&-\frac{\binom{M-F-1}{N-1}}{\binom{M-1}{N-1}}\frac{(N-1)[FVNB-(M-F-Y)rc(M-1)]}{N(M-F-1)(M-1)}-\frac{(N-1)VB}{M-1}.
\end{eqnarray*}
The payoff for pool excluders is
\begin{eqnarray*}
\Pi_{F}=rc-c-\delta.
\end{eqnarray*}
Finally, the payoff for pool punishers is
\begin{eqnarray*}
\Pi_{V}=rc-c-G-\frac{\binom{M-F-1}{N-1}}{\binom{M-1}{N-1}}\frac{rc(N-1)Y}{N(M-F-1)}.
\end{eqnarray*}

For small mutation rate, the embedded Markov chain describing the
transitions between cooperators ($X$), defectors ($Y$), pool
excluders ($F$), and pool punishers ($V$) is given by
\begin{equation*}
\left(
\begin{array}{cccc}
I_{X} & \frac{\rho_{XY}}{3} & \frac{\rho_{XF}}{3} & \frac{\rho_{XV}}{3}\\
\frac{\rho_{YX}}{3} & I_{Y} & \frac{\rho_{YF}}{3}  & \frac{\rho_{YV}}{3}\\
\frac{\rho_{FX}}{3} & \frac{\rho_{FY}}{3} & I_{F} & \frac{\rho_{FV}}{3}\\
\frac{\rho_{VX}}{3} & \frac{\rho_{VY}}{3} & \frac{\rho_{VF}}{3} &
I_{V}\\
\end{array}
\right),
\end{equation*}
where $I_{K}=1-\sum_{K\neq L} \frac{\rho_{KL}}{3}$, and $K, L\in\{X, Y, F, V\}$.

For strong imitation, we set that $\delta=G$, then the transition matrix is given by
\begin{equation*}
\left(
\begin{array}{ccccc}
\frac{2}{3} & \frac{1}{3} & 0 & 0\\
0 & \frac{2}{3} & \frac{1}{3} & 0\\
\frac{1}{3} & 0 & \frac{2}{3}-\frac{1}{3M} & \frac{1}{3M}\\
\frac{1}{3} & 0 & \frac{1}{3M} & \frac{2}{3}-\frac{1}{3M}\\
\end{array}
\right).
\end{equation*}
In this case the resulting stationary distribution is $[\frac{1}{3}
,\frac{1}{3} ,\frac{M+1}{3M+6} ,\frac{1}{3M+6}]$. It suggests that
pool punishers become extinct for large population size, and the
surviving three strategies beat each other by forming a
Rock-Paper-Scissors-like cyclic dominance [see Fig.~S1(a)].

In the presence of second-order exclusion and punishment, we assume that pool excluders exclude pool punishers, pure cooperators, and defectors. Similarly, pool punishers punish pool excluders, pure cooperators, and defectors. While the payoff for defectors remains unchanged, the average payoff for cooperators is modified by
\begin{eqnarray*}
\Pi_{X}=-c-\frac{(N-1)VB}{M-1}+\frac{\binom{M-F-1}{N-1}}{\binom{M-1}{N-1}}\{\frac{-NFVB(N-1)+rc(M-1)[N(M-F-1)-(N-1)Y]}{N(M-1)(M-F-1)}\}.
\end{eqnarray*}
The payoff for pool excluders is
\begin{eqnarray*}
\Pi_{F}=\sum_{l=0}^{N-1}\sum_{i=0}^{N-l-1}\sum_{p=0}^{N-i-l-1}\frac{\binom{M-X-Y-V-1}{l}\binom{X}{i}\binom{V}{p}\binom{Y}{N-i-p-l-1}}{\binom{M-1}{N-1}}[\frac{r(i+l+p+1)c}{l+1}-c-pB-\delta],
\end{eqnarray*}
where $l$ represents the number of pool excluders in the group. Last, the payoff for pool punishers is
\begin{eqnarray*}
\Pi_{V}=-c-G+\frac{\binom{M-F-1}{N-1}}{\binom{M-1}{N-1}}[rc-\frac{rc(N-1)Y}{N(M-F-1)}].
\end{eqnarray*}

The simplified transition matrix in the strong imitation limit is
\begin{equation*}
\left(
\begin{array}{cccc}
\frac{1}{3} & \frac{1}{3} & \frac{1}{3} & 0\\
0 & \frac{2}{3} & \frac{1}{3} & 0\\
0 & 0 & 1 & 0\\
0 & 0 & \frac{1}{3} & \frac{2}{3}\\
\end{array}
\right),
\end{equation*}
which yields the stationary distribution $[0, 0, 1, 0]$. This
distribution shows that pool excluders prevail, as shown in Fig.
S1(b). \setcounter{figure}{0}
\renewcommand{\thefigure}{S\arabic{figure}}

\begin{figure}
\begin{center}
\includegraphics[width=14cm]{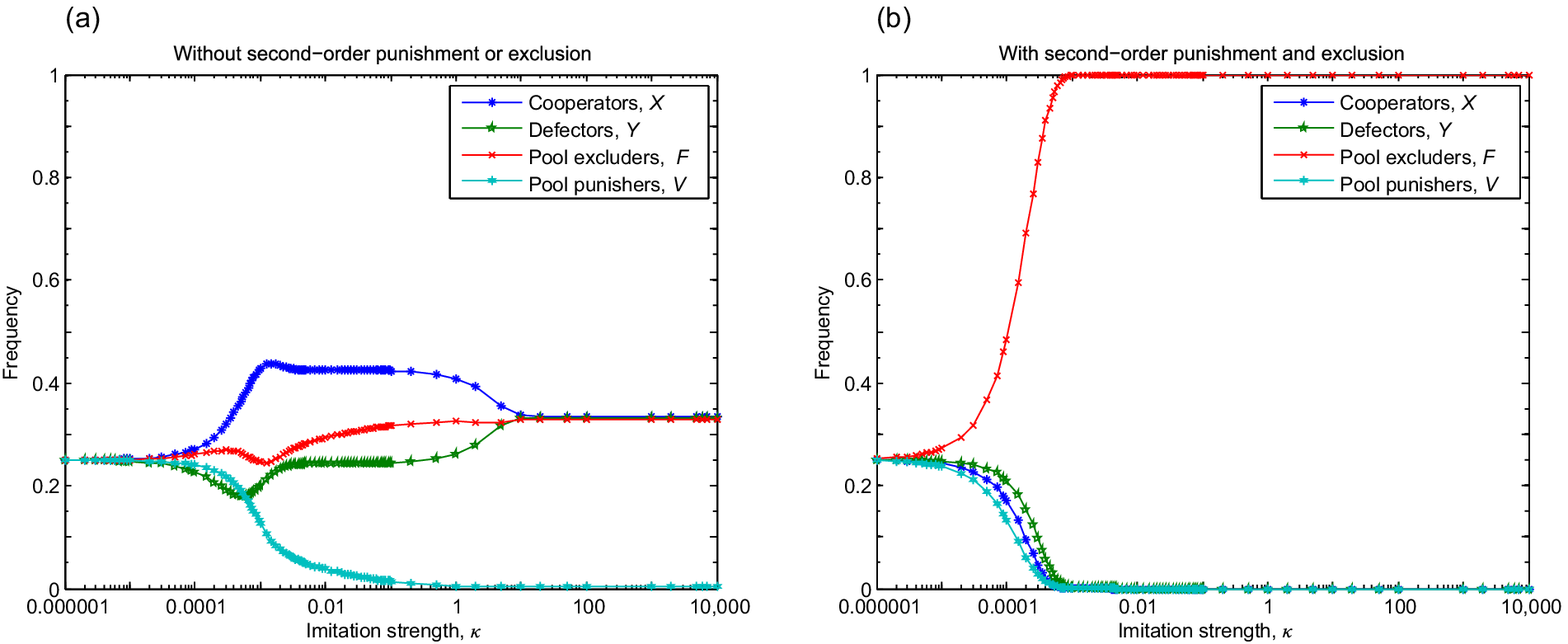} \caption{(Color
online) The competition between pool exclusion and pool punishment
in the compulsory PGG. In the absence of second-order exclusion and
punishment, punishers become extinct, and the other three strategies
can coexist in time average (a). In the presence of second-order
exclusion and punishment, pool excluders can occupy the whole
population (b). Parameters: $N=5$, $M=100$, $c=1$, $r=3$, $G=0.4$,
$\delta=0.4$, and $B=0.4$.}\label{figs1}
\end{center}
\end{figure}

\section*{3 Competition between peer exclusion and peer punishment in the optional PGG}
\noindent

In the following let us assume that there are $E$ peer excluders and
$W$ peer punishers in the population, and we have $X+Y+Z+E+W=M$. We
set that the exclusion probability is $1$, then the probability that
no peer excluders are found in the population is given by
\begin{eqnarray*}
\frac{\binom{M-E-1}{N-1}}{\binom{M-1}{N-1}}.
\end{eqnarray*}
Accordingly, in the absence of second-order exclusion and punishment the average payoff for cooperators is
\begin{eqnarray*}
\Pi_{X}&=&\frac{\binom{Z}{N-1}}{\binom{M-1}{N-1}}\sigma+[1-\frac{\binom{Z}{N-1}}{\binom{M-1}{N-1}}]\{[1-\frac{\binom{M-E-1}{N-1}}{\binom{M-1}{N-1}}](rc-c)+\\
&&\frac{\binom{M-E-1}{N-1}}{\binom{M-1}{N-1}}\sum_{i=0}^{N-1}\sum_{k=0}^{N-i-1}\frac{\binom{M-E-Y-Z-1}{i}\binom{Z}{k}\binom{Y}{N-i-k-1}}{\binom{M-E-1}{N-1}}[\frac{rc(i+1)}{N-k}-c]\}
\end{eqnarray*}
The payoff for defectors is
\begin{eqnarray*}
\Pi_{Y}&=&\frac{\binom{Z}{N-1}}{\binom{M-1}{N-1}}\sigma+[1-\frac{\binom{Z}{N-1}}{\binom{M-1}{N-1}}]\{[1-\frac{\binom{M-E-1}{N-1}}{\binom{M-1}{N-1}}][-\frac{(N-1)W\gamma}{M-1}]\\
&&+\frac{\binom{M-E-1}{N-1}}{\binom{M-1}{N-1}}\sum_{j=0}^{N-1}\sum_{t=0}^{N-1-j}\sum_{k=0}^{N-j-t-1}\frac{\binom{M-E-Y-Z-W}{N-t-k-j-1}\binom{W}{t}\binom{Z}{k}\binom{Y-1}{j}}{\binom{M-E-1}{N-1}}[\frac{rc(N-1-j-k)}{N-k}-t\gamma]\},
\end{eqnarray*}
where $t$ represents the number of peer punishers. The payoff for peer excluders is
\begin{align*}
\Pi_{E}&=\frac{\binom{Z}{N-1}}{\binom{M-1}{N-1}}\sigma+[1-\frac{\binom{Z}{N-1}}{\binom{M-1}{N-1}}][rc-c-\frac{(N-1)Yc_{E}}{M-1}].
\end{align*}
Last, peer punishers' payoff is
\begin{eqnarray*}
\Pi_{W}&=&\frac{\binom{Z}{N-1}}{\binom{M-1}{N-1}}\sigma+[1-\frac{\binom{Z}{N-1}}{\binom{M-1}{N-1}}]\{rc-c-\frac{N-1}{M-1}Y\beta-\frac{\binom{M-E-1}{N-1}}{\binom{M-1}{N-1}}[rc-c-\frac{(N-1)Y\beta}{M-1}\\
&&-\sum_{j=0}^{N-1}\sum_{k=0}^{N-j-1}\frac{\binom{M-E-Y-Z-1}{N-j-k-1}\binom{Z}{k}\binom{Y}{j}}{\binom{M-E-1}{N-1}}(\frac{N-j-k}{N-k}rc-c-j\beta)]\}.
\end{eqnarray*}

The embedded Markov chain describing the transitions between
cooperators ($X$), defectors ($Y$), loners ($Z$), peer excluders
($E$), and peer punishers ($W$) is given by
\begin{equation*}
\left(
\begin{array}{ccccc}
I_{X} & \frac{\rho_{XY}}{4} & \frac{\rho_{XZ}}{4}  & \frac{\rho_{XE}}{4} & \frac{\rho_{XW}}{4}\\
\frac{\rho_{YX}}{4} & I_{Y} & \frac{\rho_{YZ}}{4} & \frac{\rho_{YE}}{4}  & \frac{\rho_{YW}}{4}\\
\frac{\rho_{ZX}}{4} & \frac{\rho_{ZY}}{4} & I_{Z} & \frac{\rho_{ZE}}{4} & \frac{\rho_{ZW}}{4}\\
\frac{\rho_{EX}}{4} & \frac{\rho_{EY}}{4} & \frac{\rho_{EZ}}{4} & I_{E} & \frac{\rho_{EW}}{4}\\
\frac{\rho_{WX}}{4} & \frac{\rho_{WY}}{4} & \frac{\rho_{WZ}}{4} & \frac{\rho_{WE}}{4} & I_{W}\\
\end{array}
\right),
\end{equation*}
where $I_{K}=1-\sum_{K\neq L} \frac{\rho_{KL}}{4}$, and $K, L\in\{X, Y, Z, E, W\}$.

For strong imitation, the simplified transitions matrix is
\begin{equation*}
\left(
\begin{array}{ccccc}
\frac{3}{4}-\frac{1}{2M} & \frac{1}{4} & 0 & \frac{1}{4M} & \frac{1}{4M}\\
0 & \frac{1}{2} & \frac{1}{4} & \frac{1}{4} & 0\\
\frac{1}{8} & 0 & \frac{5}{8} & \frac{1}{8} & \frac{1}{8}\\
\frac{1}{4M} & 0 & 0 & 1-\frac{1}{2M} & \frac{1}{4M}\\
\frac{1}{4M} & 0 & 0 & \frac{1}{4M} & 1-\frac{1}{2M}\\
\end{array}
\right).
\end{equation*}
The resulting stationary distribution is $[\frac{6}{5M+23},
\frac{3}{5M+23}, \frac{2}{5M+23}, \frac{3M+6}{5M+23},
\frac{2M+6}{5M+23}]$, which shows that the frequency of peer
excluders is higher than the frequency of peer punishers [see
Fig.2(a)].

In the presence of second-order exclusion and punishment, we assume that peer excluders exclude peer punishers, pure cooperators, and defectors. Similarly, peer punishers punish peer excluders, pure cooperators, and defectors. The average payoff for cooperators is given by
\begin{eqnarray*}
\lefteqn{\Pi_{X}=\frac{\binom{Z}{N-1}}{\binom{M-1}{N-1}}\sigma+[1-\frac{\binom{Z}{N-1}}{\binom{M-1}{N-1}}]\{[1-\frac{\binom{M-E-1}{N-1}}{\binom{M-1}{N-1}}][-c-\frac{(N-1)W\gamma}{M-1}]+}\\
&&\frac{\binom{M-E-1}{N-1}}{\binom{M-1}{N-1}}\sum_{i=0}^{N-1}\sum_{k=0}^{N-i-1}\sum_{t=0}^{N-i-k-1}\frac{\binom{M-E-W-Y-Z-1}{i}\binom{Z}{k}\binom{W}{t}\binom{Y}{N-i-k-t-1}}{\binom{M-E-1}{N-1}}[\frac{rc(i+t+1)}{N-k}-c-t\gamma]\}.
\end{eqnarray*}
The payoff for defectors is
\begin{eqnarray*}
\Pi_{Y}&=&\frac{\binom{Z}{N-1}}{\binom{M-1}{N-1}}\sigma+[1-\frac{\binom{Z}{N-1}}{\binom{M-1}{N-1}}]\{[1-\frac{\binom{M-E-1}{N-1}}{\binom{M-1}{N-1}}][-\frac{(N-1)W\gamma}{M-1}]\\
&&+\frac{\binom{M-E-1}{N-1}}{\binom{M-1}{N-1}}\sum_{j=0}^{N-1}\sum_{t=0}^{N-1-j}\sum_{k=0}^{N-j-t-1}\frac{\binom{M-E-Y-Z-W}{N-t-k-j-1}\binom{W}{t}\binom{Z}{k}\binom{Y-1}{j}}{\binom{M-E-1}{N-1}}[\frac{rc(N-1-j-k)}{N-k}-t\gamma]\}.\end{eqnarray*}
The payoff for peer excluders is
\begin{eqnarray*}
\Pi_{E}&=&\frac{\binom{Z}{N-1}}{\binom{M-1}{N-1}}\sigma+[1-\frac{\binom{Z}{N-1}}{\binom{M-1}{N-1}}]\\
&&\sum_{i=0}^{N-1}\sum_{j=0}^{N-1-i}\sum_{k=0}^{N-i-j-1}\sum_{t=0}^{N-i-j-k-1}\frac{\binom{M-X-W-Y-Z-1}{N-i-j-k-t-1}\binom{Z}{k}\binom{X}{i}\binom{W}{t}\binom{Y}{j}}{\binom{M-1}{N-1}}[\frac{r(N-j-k)c}{N-i-j-k-t}\\
&&-c-(i+j+t)c_{E}-t\gamma].
\end{eqnarray*}
Finally, the payoff for peer punishers is
\begin{eqnarray*}
\Pi_{W}&=&\frac{\binom{Z}{N-1}}{\binom{M-1}{N-1}}\sigma+[1-\frac{\binom{Z}{N-1}}{\binom{M-1}{N-1}}]\{[1-\frac{\binom{M-E-1}{N-1}}{\binom{M-1}{N-1}}][-c-\frac{(N-1)(X+E+Y)\beta}{M-1}]+\\
&&\frac{\binom{M-E-1}{N-1}}{\binom{M-1}{N-1}}\sum_{i=0}^{N-1}\sum_{j=0}^{N-1-i}\sum_{k=0}^{N-i-j-1}\frac{\binom{M-E-X-Y-Z-1}{N-i-j-k-1}\binom{Z}{k}\binom{X}{i}\binom{Y}{j}}{\binom{M-E-1}{N-1}}[\frac{rc(N-j-k)}{N-k}-c-(i+j)\beta]\}.
\end{eqnarray*}

The transition matrix between cooperators, defectors, loners, peer excluders, and peer punishers in the strong imitation limit is
\begin{equation*}
\left(
\begin{array}{ccccc}
\frac{1}{2} & \frac{1}{4} & 0 & \frac{1}{4} & 0\\
0 & \frac{1}{2} & \frac{1}{4} & \frac{1}{4} & 0\\
\frac{1}{8} & 0 & \frac{5}{8} & \frac{1}{8} & \frac{1}{8}\\
0 & 0 & 0 & 1 & 0\\
0 & 0 & 0 & \frac{1}{4} & \frac{3}{4}\\
\end{array}
\right),
\end{equation*}
which yields the $[0, 0, 0, 1, 0]$ stationary distribution. As a
result, peer excluders prevail, shown in Fig.2(b).

\section*{4 Competition between peer exclusion and peer punishment in the compulsory PGG}
\noindent

Considering the compulsory PGG the probability that no peer excluder is found in the group is
\begin{eqnarray*}
\frac{\binom{M-E-1}{N-1}}{\binom{M-1}{N-1}}.
\end{eqnarray*}
In the absence of second-order exclusion and punishment, the average payoff for cooperators is
\begin{align*}
\Pi_{X}&=[1-\frac{\binom{M-E-1}{N-1}}{\binom{M-1}{N-1}}](rc-c)+\frac{\binom{M-E-1}{N-1}}{\binom{M-1}{N-1}}\sum_{i=0}^{N-1}\frac{\binom{M-E-Y-1}{i}\binom{Y}{N-i-1}}{\binom{M-E-1}{N-1}}[\frac{rc(i+1)}{N}-c]\nonumber\\
&=rc[1-\frac{\binom{M-E-1}{N-1}}{\binom{M-1}{N-1}}\frac{(N-1)Y}{(M-E-1)N}]-c,
\end{align*}
while the average payoff for defectors is
\begin{eqnarray*}
\Pi_{Y}=\frac{\binom{M-E-1}{N-1}}{\binom{M-1}{N-1}}\frac{rc(M-1)(N-1)(W+X)-N(N-1)WE\gamma}{N(M-E-1)(M-1)}-\frac{(N-1)W\gamma}{M-1}.
\end{eqnarray*}
The average payoff for peer excluders is
\begin{align*}
\Pi_{E}&=\sum_{i=0}^{N-1}\frac{\binom{M-Y-1}{i}\binom{Y}{N-i-1}}{\binom{M-1}{N-1}}[rc-c-(N-i-1)c_{E}]\nonumber\\
&=rc-c-\frac{(N-1)Yc_{E}}{M-1}.
\end{align*}
Last, the payoff for peer punishers is
\begin{eqnarray*}
\Pi_{W}=rc-c-\frac{(N-1)Y\beta}{M-1}-\frac{\binom{M-E-1}{N-1}}{\binom{M-1}{N-1}}\frac{rc(M-1)(N-1)Y+(N-1)NEY\beta}{N(M-E-1)(M-1)}.
\end{eqnarray*}

For small mutation rate, the embedded Markov chain describing the
transitions among cooperators ($X$), defectors ($Y$), peer excluders
($E$), and peer punishers ($W$) is given by
\begin{equation*}
\left(
\begin{array}{ccccc}
I_{X} & \frac{\rho_{XY}}{3} & \frac{\rho_{XE}}{3} & \frac{\rho_{XW}}{3}\\
\frac{\rho_{YX}}{3} & I_{Y} & \frac{\rho_{YE}}{3}  & \frac{\rho_{YW}}{3}\\
\frac{\rho_{EX}}{3} & \frac{\rho_{EY}}{3} & I_{E} & \frac{\rho_{EW}}{3}\\
\frac{\rho_{WX}}{3} & \frac{\rho_{WY}}{3} & \frac{\rho_{WE}}{3} & I_{W}\\
\end{array}
\right),
\end{equation*}
where $I_{K}=1-\sum_{K\neq L} \frac{\rho_{KL}}{3}$, and $K, L\in\{X, Y, E, W\}$.

The simplified transition matrix between cooperators ($X$),
defectors ($Y$), peer excluders ($E$), and peer punishers ($W$) in
the strong imitation can be given as
\begin{equation*}
\left(
\begin{array}{ccccc}
\frac{2}{3}-\frac{2}{3M} & \frac{1}{3} & \frac{1}{3M} & \frac{1}{3M}\\
0 & \frac{2}{3} & \frac{1}{3} & 0\\
\frac{1}{3M} & 0 & 1-\frac{2}{3M} & \frac{1}{3M}\\
\frac{1}{3M} & 0 & \frac{1}{3M} & 1-\frac{2}{3M}\\
\end{array}
\right).
\end{equation*}
The resulting stationary distribution is $[\frac{1}{4+M}, \frac{1}{4+M}, \frac{2M+3}{12+3M}, \frac{M+3}{12+3M}]$, which shows that peer excluders' frequency is higher than that of other strategists (see Fig.S2).

\begin{figure}
\begin{center}
\includegraphics[width=14cm]{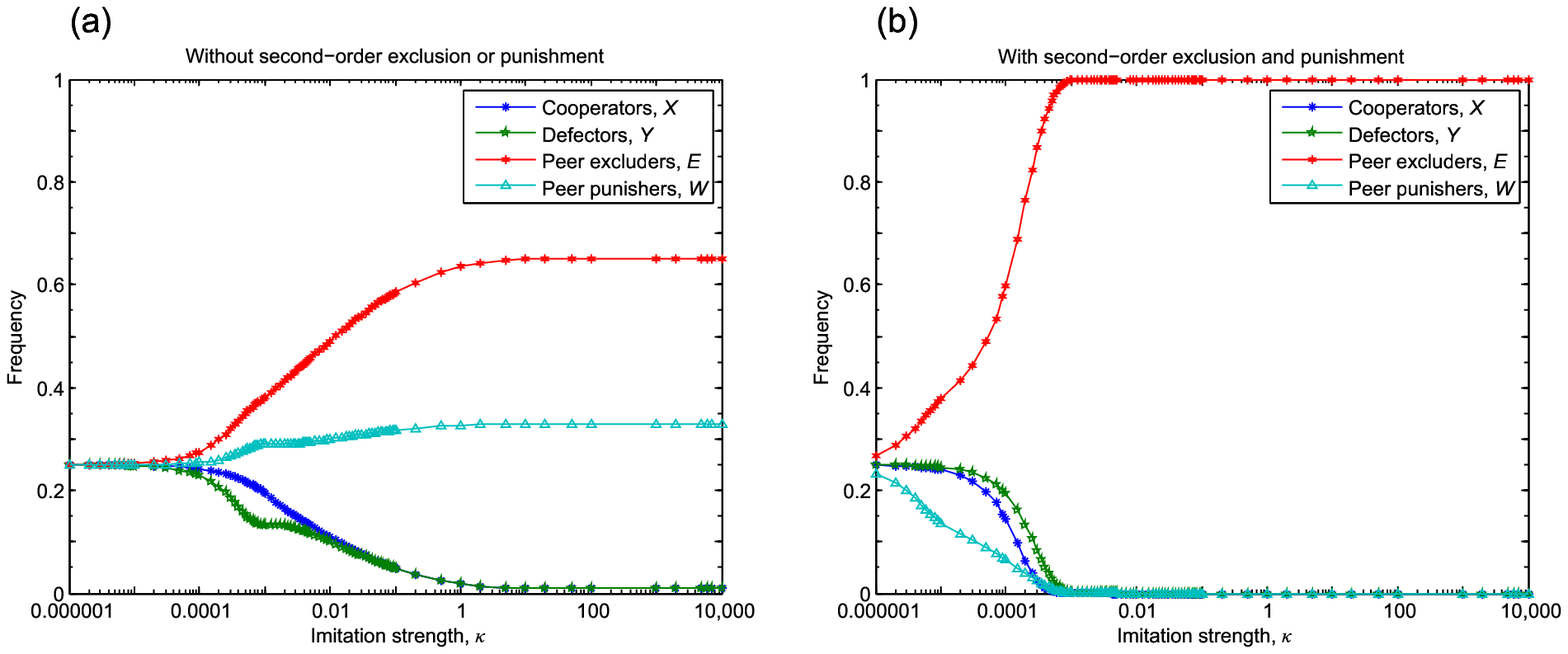}
\caption{(Color online) The competition between peer exclusion and
peer punishment in the compulsory PGG. In the absence of
second-order exclusion and punishment (a), the final frequency of
peer excluders is about two times higher than that of peer
punishers, and other two strategists become extinct (a). With
second-order exclusion and punishment, peer excluders win (b).
Parameters: $N=5$, $M=100$, $c=1$, $r=3$, $c_{E}=0.4$, $\beta=0.4$,
and $\gamma=0.4$.}\label{figs2}
\end{center}
\end{figure}

If we consider second-order exclusion and punishment, we assume that peer excluders exclude peer punishers, pure cooperators, and defectors. Simultaneously, peer punishers punish peer excluders, pure cooperators, and defectors. Accordingly, the payoff for defectors remains unchanged, but the average payoff for cooperators is changed to
\begin{align*}
\Pi_{X}=-c-\frac{(N-1)W\gamma}{M-1}+\frac{\binom{M-E-1}{N-1}}{\binom{M-1}{N-1}}\{\frac{rc(M-1)[N(M-E-1)-(N-1)Y]-NE(N-1)W\gamma}{N(M-1)(M-E-1)}\}.
\end{align*}
The payoff for peer excluders is
\begin{align*}
\Pi_{E}=\sum_{i=0}^{N-1}\sum_{j=0}^{N-1-i}\sum_{t=0}^{N-i-j-1}\frac{\binom{M-W-Y-X-1}{N-i-j-t-1}\binom{X}{i}\binom{W}{t}\binom{Y}{j}}{\binom{M-1}{N-1}}[\frac{r(N-j)c}{N-i-j-t}-c-(i+j+t)c_{E}-t\gamma],
\end{align*}
while the payoff for peer punishers is
\begin{eqnarray*}
\Pi_{W}&=&-c-\frac{(N-1)(X+E+Y)\beta}{M-1}+\frac{\binom{M-E-1}{N-1}}{\binom{M-1}{N-1}}[rc-\frac{(N-1)Yrc}{N(M-E-1)}-\\
&&\frac{(N-1)(M-W-E)\beta}{M-E-1}+\frac{(N-1)(X+E+Y)\beta}{M-1}].
\end{eqnarray*}
Here the transitions matrix between cooperators, defectors, peer excluders, and peer punishers is
\begin{equation*}
\left(
\begin{array}{ccccc}
\frac{1}{3} & \frac{1}{3} & \frac{1}{3} & 0\\
0 & \frac{2}{3} & \frac{1}{3} & 0\\
0 & 0 & 1 & 0\\
0 & 0 & \frac{1}{3} & \frac{2}{3}\\
\end{array}
\right),
\end{equation*}
which gives the stationary distribution $(0, 0, 1, 0)$. It
suggests that peer excluders prevail as shown in Fig. S2.

\section*{5 Competition between pool exclusion and peer exclusion in the optional PGG}
\noindent

In the optional PGG, the probability that no peer excluder or pool excluder is found in the group is given by
\begin{eqnarray*}
\frac{\binom{M-E-F-1}{N-1}}{\binom{M-1}{N-1}}.
\end{eqnarray*}
In the absence of second-order exclusion, the average payoff for cooperators is given by
\begin{eqnarray*}
\Pi_{X}&=&\frac{\binom{Z}{N-1}}{\binom{M-1}{N-1}}\sigma+[1-\frac{\binom{Z}{N-1}}{\binom{M-1}{N-1}}]\{[1-\frac{\binom{M-E-F-1}{N-1}}{\binom{M-1}{N-1}}](rc-c)+\\
&&\frac{\binom{M-E-F-1}{N-1}}{\binom{M-1}{N-1}}\sum_{i=0}^{N-1}\sum_{k=0}^{N-1-i}\frac{\binom{M-E-F-Y-Z-1}{i}\binom{Z}{k}\binom{Y}{N-i-k-1}}{\binom{M-E-F-1}{N-1}}[\frac{rc(i+1)}{N-k}-c]\}.
\end{eqnarray*}
The average payoff for defectors is
\begin{eqnarray*}
\Pi_{Y}=\frac{\binom{Z}{N-1}}{\binom{M-1}{N-1}}\sigma+[1-\frac{\binom{Z}{N-1}}{\binom{M-1}{N-1}}]\frac{\binom{M-E-F-1}{N-1}}{\binom{M-1}{N-1}}\sum_{i=0}^{N-1}\sum_{k=0}^{N-1-i}\frac{\binom{M-E-F-Y-Z}{i}\binom{Z}{k}\binom{Y-1}{N-i-k-1}}{\binom{M-F-E-1}{N-1}}\frac{ric}{N-k}.
\end{eqnarray*}
The average payoff for pool excluders is
\begin{eqnarray*}
\Pi_{F}=\frac{\binom{Z}{N-1}}{\binom{M-1}{N-1}}\sigma+[1-\frac{\binom{Z}{N-1}}{\binom{M-1}{N-1}}](rc-c-\delta).
\end{eqnarray*}
The average payoff for peer excluders is
\begin{eqnarray*}
\Pi_{E}=\frac{\binom{Z}{N-1}}{\binom{M-1}{N-1}}\sigma+[1-\frac{\binom{Z}{N-1}}{\binom{M-1}{N-1}}][rc-c-\frac{(N-1)Yc_{E}}{M-1}].
\end{eqnarray*}

For small mutation rate, the embedded Markov chain describing the
transitions between cooperators ($X$), defectors ($Y$), loners
($Z$), pool excluders ($F$), and peer excluders ($E$) is given by
\begin{equation*}
\left(
\begin{array}{ccccc}
I_{X} & \frac{\rho_{XY}}{4} & \frac{\rho_{XZ}}{4}  & \frac{\rho_{XF}}{4} & \frac{\rho_{XE}}{4}\\
\frac{\rho_{YX}}{4} & I_{Y} & \frac{\rho_{YZ}}{4} & \frac{\rho_{YF}}{4}  & \frac{\rho_{YE}}{4}\\
\frac{\rho_{ZX}}{4} & \frac{\rho_{ZY}}{4} & I_{Z} & \frac{\rho_{ZF}}{4} & \frac{\rho_{ZE}}{4}\\
\frac{\rho_{FX}}{4} & \frac{\rho_{FY}}{4} & \frac{\rho_{FZ}}{4} & I_{F} & \frac{\rho_{FE}}{4}\\
\frac{\rho_{EX}}{4} & \frac{\rho_{EY}}{4} & \frac{\rho_{EZ}}{4} & \frac{\rho_{EF}}{4} & I_{E}\\
\end{array}
\right),
\end{equation*}
where $I_{K}=1-\sum_{K\neq L} \frac{\rho_{KL}}{4}$, and $K, L\in\{X, Y, Z, F, E\}$.

For strong imitation, the transition matrix becomes to
\begin{equation*}
\left(
\begin{array}{ccccc}
\frac{3}{4}-\frac{1}{4M} & \frac{1}{4} & 0 & 0 & \frac{1}{4M}\\
0 & \frac{1}{4} & \frac{1}{4} & \frac{1}{4} & \frac{1}{4}\\
\frac{1}{8} & 0 & \frac{5}{8} & \frac{1}{8} & \frac{1}{8}\\
\frac{1}{4} & 0 & 0 & \frac{1}{2} & \frac{1}{4}\\
\frac{1}{4M} & 0 & 0 & 0 & 1-\frac{1}{4M}\\
\end{array}
\right).
\end{equation*}
Accordingly, the stationary distribution is $[\frac{9}{25+6M}, \frac{3}{25+6M},
\frac{2}{25+6M}, \frac{2}{25+6M}, \frac{6M+9}{25+6M}]$, and the population is dominated by peer excluders (see
Fig. 3).

If we consider second-order exclusion, we assume that pool excluders
exclude peer excluders because the latter players do not contribute
to the exclusion pool. But peer excluders do not exclude pool
excluders. Accordingly, the average payoff for cooperators is given
by
\begin{eqnarray*}
\Pi_{X}&=&\frac{\binom{Z}{N-1}}{\binom{M-1}{N-1}}\sigma+[1-\frac{\binom{Z}{N-1}}{\binom{M-1}{N-1}}]\{[1-\frac{\binom{M-E-F-1}{N-1}}{\binom{M-1}{N-1}}](-c)+\\
&&\frac{\binom{M-E-F-1}{N-1}}{\binom{M-1}{N-1}}\sum_{i=0}^{N-1}\sum_{k=0}^{N-1-i}\frac{\binom{M-E-F-Y-Z-1}{i}\binom{Z}{k}\binom{Y}{N-i-k-1}}{\binom{M-E-F-1}{N-1}}[\frac{rc(i+1)}{N-k}-c]\},
\end{eqnarray*}
where $\frac{\binom{M-E-F-1}{N-1}}{\binom{M-1}{N-1}}$ denotes the
probability that neither peer excluder nor pool excluder is found in
the group. The average payoffs for peer excluders and pool excluders
are respectively
\begin{eqnarray*}
\Pi_{E}&=&\frac{\binom{Z}{N-1}}{\binom{M-1}{N-1}}\sigma+[1-\frac{\binom{Z}{N-1}}{\binom{M-1}{N-1}}]\{[1-\frac{\binom{M-F-1}{N-1}}{\binom{M-1}{N-1}}][-c-\sum_{i=0}^{N-1}\sum_{j=0}^{N-1-i}\frac{\binom{M-Y-X-1}{N-i-j-1}\binom{X}{i}\binom{Y}{j}}{\binom{M-1}{N-1}}(i+j)c_{E}]\\
&&+\frac{\binom{M-F-1}{N-1}}{\binom{M-1}{N-1}}\sum_{i=0}^{N-1}\sum_{k=0}^{N-1-i}\sum_{j=0}^{N-1-i-k}\frac{\binom{M-X-F-Y-Z-1}{N-i-j-k-1}\binom{Z}{k}\binom{Y}{j}\binom{X}{i}}{\binom{M-F-1}{N-1}}[\frac{rc(N-j-k)}{N-j-i-k}-c-(i+j)c_{E}]\},
\end{eqnarray*}
\begin{eqnarray*}
\Pi_{F}=\frac{\binom{Z}{N-1}}{\binom{M-1}{N-1}}\sigma+[1-\frac{\binom{Z}{N-1}}{\binom{M-1}{N-1}}]\sum_{k=0}^{N-1}\sum_{j=0}^{N-1-k}\sum_{l=0}^{N-1-k-j}\frac{\binom{M-F-Y-Z}{N-j-k-l-1}\binom{Z}{k}\binom{Y}{j}\binom{F-1}{l}}{\binom{M-1}{N-1}}[\frac{rc(N-j-k)}{l+1}-c-\delta].
\end{eqnarray*}
The payoffs of loners and defectors are not changed.

For strong imitation, the transitions between cooperators ($X$),
defectors ($Y$), loners ($Z$), pool excluders ($F$), and peer
excluders ($E$) are given by
\begin{equation*}
\left(
\begin{array}{ccccc}
\frac{1}{4} & \frac{1}{4} & 0 & \frac{1}{4} & \frac{1}{4}\\
0 & \frac{1}{4} & \frac{1}{4} & \frac{1}{4} & \frac{1}{4}\\
\frac{1}{8} & 0 & \frac{5}{8} & \frac{1}{8} & \frac{1}{8}\\
0 & 0 & 0 & 1 & 0\\
0 & 0 & 0 & \frac{1}{4} & \frac{3}{4}\\
\end{array}
\right).
\end{equation*}
Accordingly, the stationary distribution is $[0, 0, 0, 1, 0]$, which
shows that pool excluders prevail (see Fig. 3).

\section*{6 Competition between pool exclusion and peer exclusion in the compulsory PGG}
\noindent

In the compulsory PGG, the probability that no peer excluder or no pool excluder is found in the group is given by
\begin{eqnarray*}
\frac{\binom{M-E-F-1}{N-1}}{\binom{M-1}{N-1}}.
\end{eqnarray*}
In the absence of second-order exclusion, the average payoff for cooperators is given by
\begin{align*}
\Pi_{X}&=[1-\frac{\binom{M-E-F-1}{N-1}}{\binom{M-1}{N-1}}](rc-c)+\frac{\binom{M-E-F-1}{N-1}}{\binom{M-1}{N-1}}\sum_{i=0}^{N-1}\frac{\binom{M-E-F-Y-1}{i}\binom{Y}{N-i-1}}{\binom{M-E-F-1}{N-1}}[\frac{rc(i+1)}{N}-c]\nonumber\\
&=rc[1-\frac{\binom{M-E-F-1}{N-1}}{\binom{M-1}{N-1}}\frac{(N-1)Y}{N(M-E-F-1)}]-c.
\end{align*}
The payoff for defectors is
\begin{align*}
\Pi_{Y}&=\frac{\binom{M-E-F-1}{N-1}}{\binom{M-1}{N-1}}\sum_{i=0}^{N-1}\frac{\binom{M-E-F-Y}{i}\binom{Y-1}{N-i-1}}{\binom{M-E-F-1}{N-1}}\frac{rci}{N}\nonumber\\
&=\frac{\binom{M-E-F-1}{N-1}}{\binom{M-1}{N-1}}\frac{rc}{N}\frac{(N-1)(M-E-F-Y)}{M-E-F-1},
\end{align*}
while the average payoff for pool excluders is
\begin{eqnarray*}
\Pi_{F}=rc-c-\delta.
\end{eqnarray*}
Last, the average payoff for peer excluders is
\begin{eqnarray*}
\Pi_{E}=rc-c-\frac{(N-1)Yc_{E}}{M-1}.
\end{eqnarray*}

For small mutation rate, the embedded Markov chain describing the
transitions between cooperators ($X$), defectors ($Y$), pool
excluders ($F$), and peer excluders ($E$) is given by
\begin{equation*}
\left(
\begin{array}{cccc}
I_{X} & \frac{\rho_{XY}}{3} & \frac{\rho_{XF}}{3} & \frac{\rho_{XE}}{3}\\
\frac{\rho_{YX}}{3} & I_{Y} & \frac{\rho_{YF}}{3}  & \frac{\rho_{YE}}{3}\\
\frac{\rho_{FX}}{3} & \frac{\rho_{FY}}{3} & I_{F} & \frac{\rho_{FE}}{3}\\
\frac{\rho_{EX}}{3} & \frac{\rho_{EY}}{3} & \frac{\rho_{EF}}{3} & I_{E}\\
\end{array}
\right),
\end{equation*}
where $I_{K}=1-\sum_{K\neq L} \frac{\rho_{KL}}{3}$, and $K, L\in\{X, Y, F, E\}$.

For strong imitation, the transition matrix becomes
\begin{equation*}
\left(
\begin{array}{ccccc}
\frac{2}{3}-\frac{1}{3M} & \frac{1}{3} & 0 & \frac{1}{3M} \\
0 & \frac{1}{3} & \frac{1}{3} & \frac{1}{3}\\
\frac{1}{3} & 0 & \frac{1}{3} & \frac{1}{3}\\
\frac{1}{3M} & 0 & 0 & 1-\frac{1}{3M}\\
\end{array}
\right).
\end{equation*}
The resulting stationary distribution is $ [\frac{4}{3M+11}, \frac{2}{3M+11},
\frac{1}{3M+11}, \frac{3M+4}{3M+11}]$, which suggests that the population is dominated by peer
excluders (see Fig.S3).

\begin{figure}
\begin{center}
\includegraphics[width=14cm]{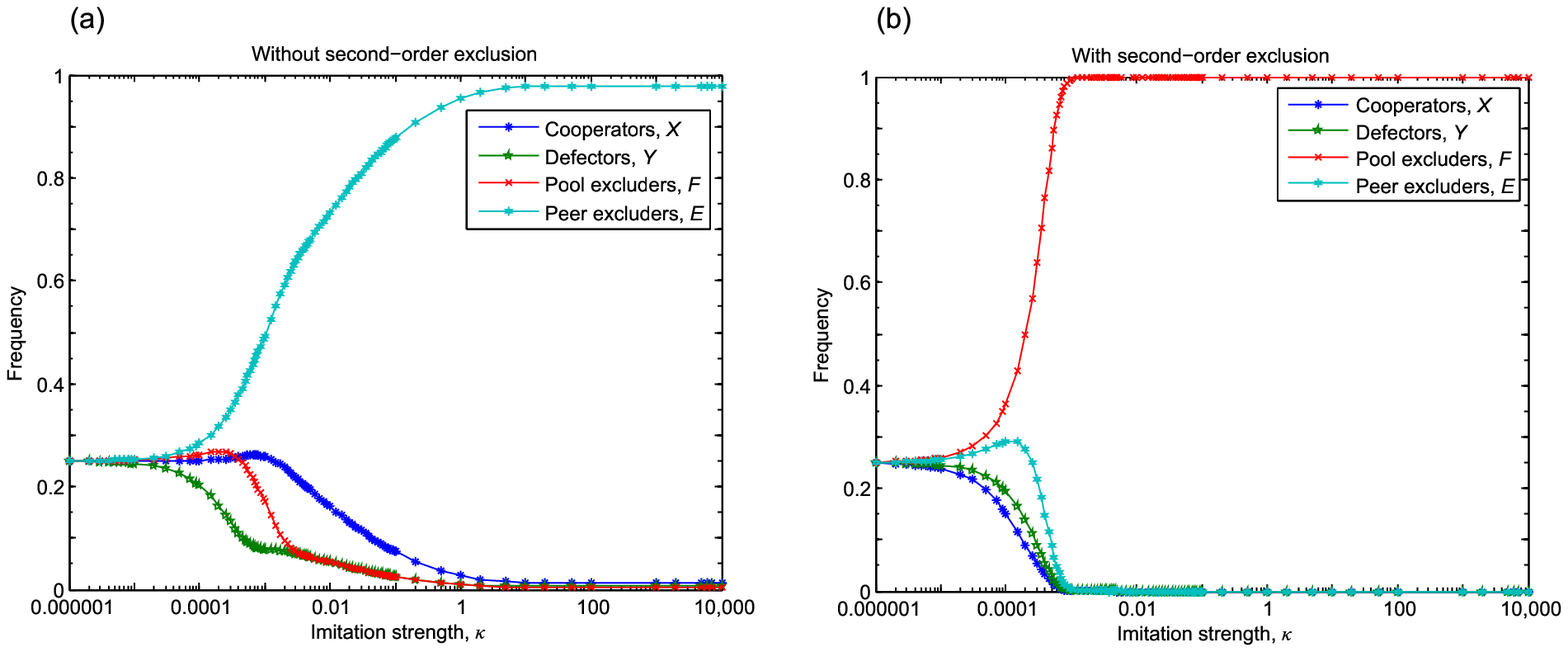}
\caption{(Color online) The competition between peer and pool
exclusion in the compulsory PGG. In the absence of second-order
exclusion, peer excluders can occupy the whole population, and other
strategists become extinct (a). With second-order exclusion, pool
excluders win (b). Parameters: $N=5$, $M=100$, $c=1$, $r=3$,
$c_{E}=0.4$, and $\delta=0.4$.} \label{figs3}
\end{center}
\end{figure}

In the presence of second-order exclusion, we assume that pool excluders exclude peer excluders, but peer excluders do not exclude pool excluders. The average payoff for cooperators is given by
\begin{align*}
\Pi_{X}&=[1-\frac{\binom{M-E-F-1}{N-1}}{\binom{M-1}{N-1}}](-c)+\frac{\binom{M-E-F-1}{N-1}}{\binom{M-1}{N-1}}\sum_{i=0}^{N-1}\frac{\binom{M-E-F-Y-1}{i}\binom{Y}{N-i-1}}{\binom{M-E-F-1}{N-1}}[\frac{rc(i+1)}{N}-c]\nonumber\\
&=\frac{\binom{M-E-F-1}{N-1}}{\binom{M-1}{N-1}}\{\frac{rc}{N}[\frac{(M-Y-E-F-1)(N-1)}{M-E-F-1}+1]\}-c.
\end{align*}
The average payoffs for pool excluders and peer excluders are
respectively given by
\begin{eqnarray*}
\Pi_{F}&=&\sum_{l=0}^{N-1}\sum_{i=0}^{N-l-1}\frac{\binom{M-X-E-Y-1}{l}\binom{Y}{N-i-l-1}\binom{X+E}{i}}{\binom{M-1}{N-1}}[\frac{rc(i+l+1)}{l+1}-c-\delta],\\
\Pi_{E}&=&[1-\frac{\binom{M-F-1}{N-1}}{\binom{M-1}{N-1}}][-c-\frac{(N-1)(X+Y)}{M-1}c_{E}]+\\
&&\frac{\binom{M-F-1}{N-1}}{\binom{M-1}{N-1}}\sum_{i=0}^{N-1}\sum_{j=0}^{N-i-1}\frac{\binom{M-X-F-Y-1}{N-j-i-1}\binom{Y}{j}\binom{X}{i}}{\binom{M-F-1}{N-1}}[\frac{rc(N-j)}{N-i-j}-c-(i+j)c_{E}].
\end{eqnarray*}

For strong imitation, the transition matrix is
\begin{equation*}
\left(
\begin{array}{ccccc}
0 & \frac{1}{3} & \frac{1}{3} & \frac{1}{3}\\
0 & \frac{1}{3} & \frac{1}{3} & \frac{1}{3}\\
0 & 0 & 1 & 0\\
0 & 0 & \frac{1}{3} & \frac{2}{3}\\
\end{array}
\right).
\end{equation*}
The resulting stationary distribution is $[0, 0, 1,0]$, namely pool
excluders prevail (see Fig. S3).

\section*{7 Competition between prosocial exclusions and punishments in the optional PGG}
\noindent

In this case the population contains $X$ cooperators, $Y$ defectors,
$Z$ loners, $E$ peer excluders, $W$ peer punishers, $F$ pool
excluders, and $V$ pool punishers whose fractions fulfill the
constraint $X+Y+Z+E+W+F+V=M$. In the absence of second-order
exclusion and punishment, the average payoff for cooperators is
\begin{eqnarray*}
\Pi_{X}&=&\frac{\binom{Z}{N-1}}{\binom{M-1}{N-1}}\sigma+[1-\frac{\binom{Z}{N-1}}{\binom{M-1}{N-1}}]\{[1-\frac{\binom{M-E-F-1}{N-1}}{\binom{M-1}{N-1}}](rc-c)+\\
&&\frac{\binom{M-E-F-1}{N-1}}{\binom{M-1}{N-1}}\sum_{i=0}^{N-1}\sum_{k=0}^{N-1-i}\frac{\binom{M-E-F-Y-Z-1}{i}\binom{Z}{k}\binom{Y}{N-i-k-1}}{\binom{M-E-F-1}{N-1}}[\frac{rc(i+1)}{N-k}-c]\}.
\end{eqnarray*}
The payoff for defectors is
\begin{eqnarray*}
\lefteqn{\Pi_{Y}=\frac{\binom{Z}{N-1}}{\binom{M-1}{N-1}}\sigma+[1-\frac{\binom{Z}{N-1}}{\binom{M-1}{N-1}}]\{[1-\frac{\binom{M-E-F-1}{N-1}}{\binom{M-1}{N-1}}][-\frac{(N-1)W\gamma+(N-1)VB}{M-1}]+\frac{\binom{M-E-F-1}{N-1}}{\binom{M-1}{N-1}}}\\
&&\sum_{j=0}^{N-1}\sum_{t=0}^{N-1-j}\sum_{k=0}^{N-1-j-t}\sum_{p=0}^{N-1-j-t-k}\frac{\binom{M-E-F-Y-Z-W-V}{N-j-t-k-p-1}\binom{W}{t}\binom{Z}{k}\binom{V}{p}\binom{Y-1}{j}}{\binom{M-F-E-1}{N-1}}[\frac{r(N-j-k-1)c}{N-k}-t\gamma-pB]\}.
\end{eqnarray*}
The average payoff for peer excluders is
\begin{eqnarray*}
\Pi_{E}=\frac{\binom{Z}{N-1}}{\binom{M-1}{N-1}}\sigma+[1-\frac{\binom{Z}{N-1}}{\binom{M-1}{N-1}}][rc-c-\frac{(N-1)Yc_{E}}{M-1}].
\end{eqnarray*}
The average payoff for peer punishers is
\begin{eqnarray*}
\Pi_{W}&=&\frac{\binom{Z}{N-1}}{\binom{M-1}{N-1}}\sigma+[1-\frac{\binom{Z}{N-1}}{\binom{M-1}{N-1}}]\{[1-\frac{\binom{M-E-F-1}{N-1}}{\binom{M-1}{N-1}}][rc-c-\frac{(N-1)Y\beta}{M-1}]\\
&&+\frac{\binom{M-E-F-1}{N-1}}{\binom{M-1}{N-1}}\sum_{i=0}^{N-1}\sum_{k=0}^{N-1-i}\frac{\binom{M-E-F-Y-Z-1}{i}\binom{Z}{k}\binom{Y}{N-i-k-1}}{\binom{M-F-E-1}{N-1}}[\frac{rc(i+1)}{N-k}-c-(N-i-k-1)\beta]\}.
\end{eqnarray*}
The average payoff for pool excluders is
\begin{eqnarray*}
\Pi_{F}=\frac{\binom{Z}{N-1}}{\binom{M-1}{N-1}}\sigma+[1-\frac{\binom{Z}{N-1}}{\binom{M-1}{N-1}}](rc-c-\delta).
\end{eqnarray*}
Last, the average payoff for pool punishers is
\begin{eqnarray*}
\Pi_{V}&=&\frac{\binom{Z}{N-1}}{\binom{M-1}{N-1}}\sigma+[1-\frac{\binom{Z}{N-1}}{\binom{M-1}{N-1}}]\{[1-\frac{\binom{M-E-F-1}{N-1}}{\binom{M-1}{N-1}}](rc-c-G)\\
&&+\frac{\binom{M-E-F-1}{N-1}}{\binom{M-1}{N-1}}\sum_{j=0}^{N-1}\sum_{k=0}^{N-j-1}\frac{\binom{M-E-F-Y-Z-1}{N-j-k-1}\binom{Y}{j}\binom{Z}{k}}{\binom{M-E-F-1}{N-1}}[\frac{r(N-j-k)c}{N-k}-c-G]\}.
\end{eqnarray*}

For small exploration rates, the embedded Markov chain describing
the transitions between cooperators ($X$), defectors ($Y$), loners
($Z$), peer excluders ($E$), peer punishers ($W$), pool excluders
($F$), and pool punishers ($V$) is given by
\begin{equation*}
\left(
\begin{array}{ccccccc}
I_{X} & \frac{\rho_{XY}}{6} & \frac{\rho_{XZ}}{6}  & \frac{\rho_{XE}}{6} & \frac{\rho_{XW}}{6} & \frac{\rho_{XF}}{6} & \frac{\rho_{XV}}{6}\\
\frac{\rho_{YX}}{6} & I_{Y} & \frac{\rho_{YZ}}{6} & \frac{\rho_{YE}}{6}  & \frac{\rho_{YW}}{6} & \frac{\rho_{YF}}{6} & \frac{\rho_{YV}}{6}\\
\frac{\rho_{ZX}}{6} & \frac{\rho_{ZY}}{6} & I_{Z} & \frac{\rho_{ZE}}{6} & \frac{\rho_{ZW}}{6} & \frac{\rho_{ZF}}{6} & \frac{\rho_{ZV}}{6}\\
\frac{\rho_{EX}}{6} & \frac{\rho_{EY}}{6} & \frac{\rho_{EZ}}{6} & I_{E} & \frac{\rho_{EW}}{6} & \frac{\rho_{EF}}{6} & \frac{\rho_{EV}}{6}\\
\frac{\rho_{WX}}{6} & \frac{\rho_{WY}}{6} & \frac{\rho_{WZ}}{6} & \frac{\rho_{WE}}{6} & I_{W} & \frac{\rho_{WF}}{6} & \frac{\rho_{WV}}{6}\\
\frac{\rho_{FX}}{6} & \frac{\rho_{FY}}{6} & \frac{\rho_{FZ}}{6} & \frac{\rho_{FE}}{6} & \frac{\rho_{FW}}{6} & I_{F} & \frac{\rho_{FV}}{6}\\
\frac{\rho_{VX}}{6} & \frac{\rho_{VY}}{6} & \frac{\rho_{VZ}}{6} & \frac{\rho_{VE}}{6} & \frac{\rho_{VW}}{6} & \frac{\rho_{VF}}{6} & I_{V}\\
\end{array}
\right),
\end{equation*}
where $I_{K}=1-\sum_{K\neq L} \frac{\rho_{KL}}{6}$, and $K, L\in\{X, Y, Z, E, W, F, V\}$.

For strong imitation, the transitions matrix is
\begin{equation*}
\left(
\begin{array}{ccccccc}
\frac{5}{6}-\frac{1}{3M} & \frac{1}{6} & 0 & \frac{1}{6M} & \frac{1}{6M} & 0 & 0\\
0 & \frac{1}{2} & \frac{1}{6} & \frac{1}{6}  & 0 & \frac{1}{6} & 0\\
\frac{1}{12} & 0 & \frac{7}{12} & \frac{1}{12} & \frac{1}{12} & \frac{1}{12} & \frac{1}{12}\\
\frac{1}{6M} & 0 & 0 & 1-\frac{1}{3M} & \frac{1}{6M} & 0 & 0\\
\frac{1}{6M} & 0 & 0 & \frac{1}{6M} & 1-\frac{1}{3M} & 0 & 0\\
\frac{1}{6} & 0 & 0 & \frac{1}{6} & \frac{1}{6} & \frac{1}{2}-\frac{1}{6M} & \frac{1}{6M}\\
\frac{1}{6} & 0 & 0 & \frac{1}{6} & \frac{1}{6} & \frac{1}{6M} & \frac{1}{2}-\frac{1}{6M}\\
\end{array}
\right).
\end{equation*}
Accordingly, the stationary distribution is $[\frac{45}{163+35M},
\frac{15}{163+35M}, \frac{6}{163+35M},  \frac{20M+45}{163+35M},
\frac{15M+45}{163+35M},\\ \frac{18M+7}{(163+35M)(3M+2)},
\frac{3M+7}{(163+35M)(3M+2)}]$, suggesting that the frequency of
peer excluders is higher than the fraction of any other strategy
(see Fig.4).

If we consider the second-order exclusion and punishment, then the average payoff for cooperators is given by
\begin{eqnarray*}
\lefteqn{\Pi_{X}=\frac{\binom{Z}{N-1}}{\binom{M-1}{N-1}}\sigma+[1-\frac{\binom{Z}{N-1}}{\binom{M-1}{N-1}}]\{[1-\frac{\binom{M-E-F-1}{N-1}}{\binom{M-1}{N-1}}][-c-\frac{(N-1)W\gamma}{M-1}-\frac{(N-1)VB}{M-1}]+}\\
&&\frac{\binom{M-E-F-1}{N-1}}{\binom{M-1}{N-1}}\sum_{i=0}^{N-1}\sum_{t=0}^{N-1-i}\sum_{p=0}^{N-1-i-t}\sum_{k=0}^{N-1-i-t-p}\frac{\binom{M-E-F-V-Y-W-Z-1}{i}\binom{Y}{N-i-t-p-k-1}\binom{W}{t}\binom{V}{p}\binom{Z}{k}}{\binom{M-E-F-1}{N-1}}\\
&&\times[\frac{rc(i+t+p+1)}{N-k}-c-t\gamma-pB]\}.
\end{eqnarray*}
The payoff for peer excluders is
\begin{eqnarray*}
\lefteqn{\Pi_{E}=\frac{\binom{Z}{N-1}}{\binom{M-1}{N-1}}\sigma+[1-\frac{\binom{Z}{N-1}}{\binom{M-1}{N-1}}]\{[1-\frac{\binom{M-F-1}{N-1}}{\binom{M-1}{N-1}}][-c-}\\
&&\frac{W\gamma+VB+(X+Y+W+V)c_{E}}{M-1}(N-1)]+\frac{\binom{M-F-1}{N-1}}{\binom{M-1}{N-1}}\\
&&\sum_{i=0}^{N-1}\sum_{j=0}^{N-1-i}\sum_{k=0}^{N-1-i-j}\sum_{p=0}^{N-1-i-j-k}\sum_{t=0}^{N-1-i-j-k-p}\frac{\binom{M-F-X-W-Y-Z-V-1}{N-i-j-k-p-t-1}\binom{X}{i}\binom{Y}{j}\binom{Z}{k}\binom{W}{t}\binom{V}{p}}{\binom{M-F-1}{N-1}}\nonumber\\
&&\times[\frac{rc(N-j-k)}{N-i-j-k-p-t}-t\gamma-pB-(i+j+t+p)c_{E}-c]\}.\nonumber\\
\end{eqnarray*}
The average payoff for peer punishers is
\begin{eqnarray*}
\lefteqn{\Pi_{W}=\frac{\binom{Z}{N-1}}{\binom{M-1}{N-1}}\sigma+[1-\frac{\binom{Z}{N-1}}{\binom{M-1}{N-1}}]\{[1-\frac{\binom{M-E-F-1}{N-1}}{\binom{M-1}{N-1}}][-c-}\nonumber\\
&&(N-1)\frac{VB+(X+Y+E+F)\beta}{M-1}]+\frac{\binom{M-E-F-1}{N-1}}{\binom{M-1}{N-1}}\nonumber\\
&&\sum_{i=0}^{N-1}\sum_{j=0}^{N-i-1}\sum_{k=0}^{N-i-j-1}\sum_{p=0}^{N-i-j-k-1}\frac{\binom{M-F-X-Z-V-Y-E-1}{N-i-j-p-k-1}\binom{V}{p}\binom{X}{i}\binom{Z}{k}\binom{Y}{j}}{\binom{M-E-F-1}{N-1}}\nonumber\\
&&\times[\frac{rc(N-j-k)}{N-k}-c-(i+j)\beta-pB]\}.
\end{eqnarray*}
The average payoff for pool excluders is
\begin{eqnarray*}
\lefteqn{\Pi_{F}=\frac{\binom{Z}{N-1}}{\binom{M-1}{N-1}}\sigma+[1-\frac{\binom{Z}{N-1}}{\binom{M-1}{N-1}}]}\nonumber\\
&&\sum_{j=0}^{N-1}\sum_{t=0}^{N-j-1}\sum_{p=0}^{N-t-j-1}\sum_{l=0}^{N-t-j-p-1}\frac{\binom{M-F-Z-V-Y-W}{N-t-j-p-l-1}\binom{V}{p}\binom{Y+Z}{j}\binom{W}{t}\binom{F-1}{l}}{\binom{M-1}{N-1}}\nonumber\\
&&\times[\frac{rc(N-j)}{l+1}-c-t\gamma-pB-\delta]\}.
\end{eqnarray*}
Last, the payoff for pool punishers is
\begin{eqnarray*}
\Pi_{V}&=&\frac{\binom{Z}{N-1}}{\binom{M-1}{N-1}}\sigma+[1-\frac{\binom{Z}{N-1}}{\binom{M-1}{N-1}}]\{[1-\frac{\binom{M-E-F-1}{N-1}}{\binom{M-1}{N-1}}](-c-G)+\frac{\binom{M-E-F-1}{N-1}}{\binom{M-1}{N-1}}\nonumber\\
&&\sum_{j=0}^{N-1}\sum_{k=0}^{N-j-1}\frac{\binom{Y}{j}\binom{Z}{k}\binom{M-E-F-Y-Z-1}{N-j-k-1}}{\binom{M-E-F-1}{N-1}}[\frac{r(N-j-k)c}{N-k}-c-G]\}.
\end{eqnarray*}

For strong imitation limit, the embedded Markov chain describing the
transition matrix between cooperators ($X$), defectors ($Y$), loners
($Z$), peer excluders ($E$), peer punishers ($W$), pool excluders
($F$), and pool punishers ($V$) is
\begin{equation}
\left(
\begin{array}{ccccccc}
\frac{1}{2} & \frac{1}{6} & 0 & \frac{1}{6} & 0 & \frac{1}{6} & 0\\
0 & \frac{1}{2} & \frac{1}{6} & \frac{1}{6} & 0 & \frac{1}{6} & 0\\
\frac{1}{12} & 0 & \frac{7}{12} & \frac{1}{12} & \frac{1}{12} & \frac{1}{12} & \frac{1}{12}\\
0 & 0 & 0 & \frac{5}{6} & 0 & \frac{1}{6} & 0\\
0 & 0 & 0 & \frac{1}{6} & \frac{2}{3} & \frac{1}{6} & 0\\
0 & 0 & 0 & 0 & 0 & 1 & 0\\
0 & 0 & 0 & \frac{1}{6} & 0 & \frac{1}{6} & \frac{2}{3}\\
\end{array}
\right).
\end{equation}
The resulting stationary distribution $[0, 0, 0, 0, 0, 1, 0]$
suggests that pool excluders prevail (see Fig.~$4$).

\section*{8 Competition between prosocial exclusions and punishments in the compulsory PGG}
\noindent

For the compulsory PGG, in the absence of second-order exclusion and
punishment the average payoff for cooperators is given by
\begin{align*}
\Pi_{X}=rc-c-\frac{\binom{M-E-F-1}{N-1}}{\binom{M-1}{N-1}}\frac{rc(N-1)Y}{N(M-E-F-1)}.
\end{align*}
The payoff for defectors is
\begin{eqnarray*}
\Pi_{Y}&=&\frac{(N-1)(-W\gamma-VB)}{M-1}-\frac{\binom{M-E-F-1}{N-1}}{\binom{M-1}{N-1}}[\frac{(N-1)(-W\gamma-VB)}{M-1}-\\
&&\frac{rc(N-1)(M-E-F-Y)-N(N-1)(W\gamma+VB)}{N(M-E-F-1)}].
\end{eqnarray*}
The payoff for peer excluders is
\begin{align*}
\Pi_{E}=rc-c-\frac{(N-1)Yc_{E}}{M-1}.
\end{align*}
The payoff for peer punishers is
\begin{align*}
\Pi_{W}=rc-c-\frac{(N-1)Y\beta}{M-1}-\frac{\binom{M-E-F-1}{N-1}}{\binom{M-1}{N-1}}[\frac{rc(N-1)(M-1)Y-N(E+F)(N-1)Y\beta}{N(M-1)(M-E-F-1)}].
\end{align*}
The payoff for pool excluders is
\begin{align*}
\Pi_{F}=rc-c-\delta.
\end{align*}
The payoff for pool punishers is
\begin{align*}
\Pi_{V}=rc-c-G-\frac{\binom{M-E-F-1}{N-1}}{\binom{M-1}{N-1}}\frac{rc}{N}\frac{(N-1)Y}{M-E-F-1}.
\end{align*}

\begin{figure}
\begin{center}
\includegraphics[width=14cm]{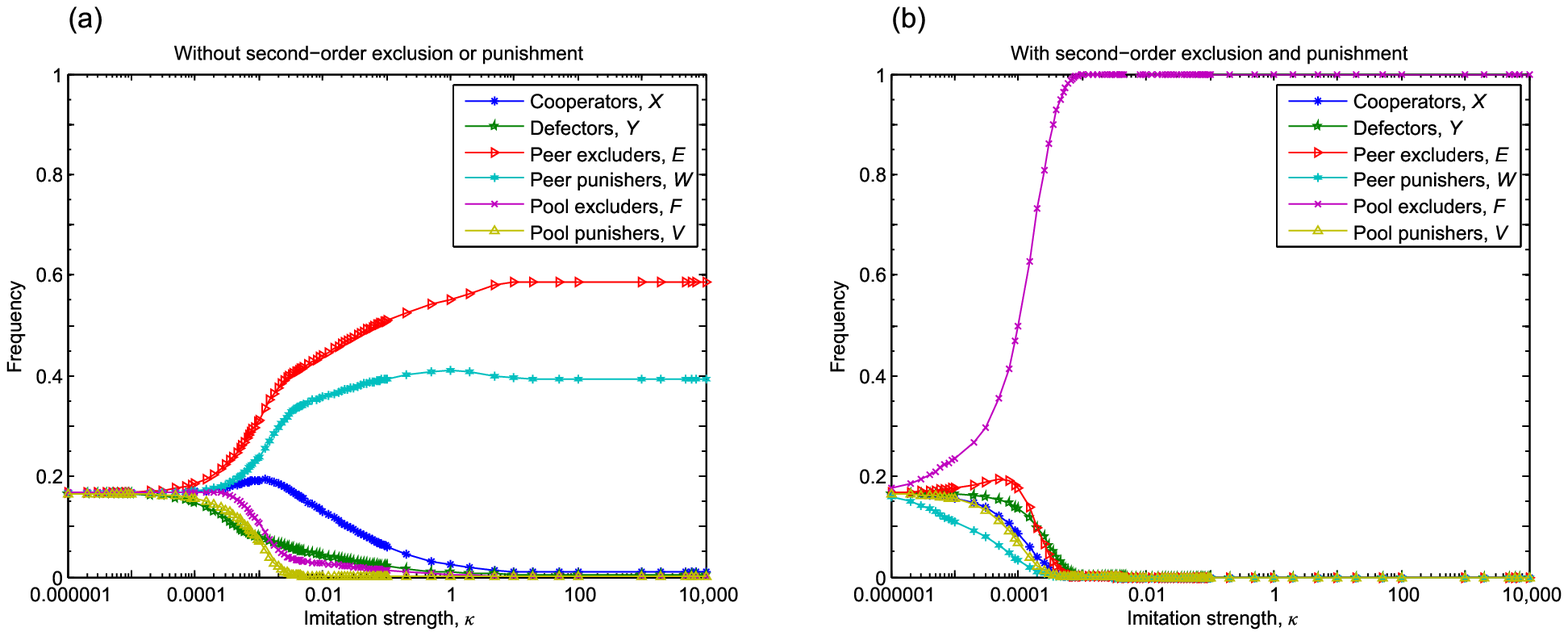}
\caption{(Color online) The competition among peer excluders, pool
excluders, peer punishers, and pool punishers in the compulsory PGG.
In the absence of second-order exclusion and punishment, peer
excluders can outperform other strategies, and peer punishers can
survive in the population (a). With second-order exclusion and
punishment, pool excluders win (b). Parameters: $N=5$, $M=100$,
$c=1$, $r=3$, $c_{E}=0.4$, $\delta=0.4$, $B=G=0.4$, and
$\beta=\gamma=0.4$.} \label{figs4}
\end{center}
\end{figure}

For small exploration rates, the transition matrix between
cooperators ($X$), defectors ($Y$), peer excluders ($E$), peer
punishers ($W$), pool excluders ($F$), and pool punishers ($V$) is
given by
\begin{equation*}
\left(
\begin{array}{cccccc}
I_{X} & \frac{\rho_{XY}}{5} & \frac{\rho_{XE}}{5} & \frac{\rho_{XW}}{5} & \frac{\rho_{XF}}{5} & \frac{\rho_{XV}}{5}\\
\frac{\rho_{YX}}{5} & I_{Y} & \frac{\rho_{YE}}{5}  & \frac{\rho_{YW}}{5} & \frac{\rho_{YF}}{5} & \frac{\rho_{YV}}{5}\\
\frac{\rho_{EX}}{5} & \frac{\rho_{EY}}{5} & I_{E} & \frac{\rho_{EW}}{5} & \frac{\rho_{EF}}{5} & \frac{\rho_{EV}}{5}\\
\frac{\rho_{WX}}{5} & \frac{\rho_{WY}}{5} & \frac{\rho_{WE}}{5} & I_{W} & \frac{\rho_{WF}}{5} & \frac{\rho_{WV}}{5}\\
\frac{\rho_{FX}}{5} & \frac{\rho_{FY}}{5} & \frac{\rho_{FE}}{5} & \frac{\rho_{FW}}{5} & I_{F} & \frac{\rho_{FV}}{5}\\
\frac{\rho_{VX}}{5} & \frac{\rho_{VY}}{5} & \frac{\rho_{VE}}{5} & \frac{\rho_{VW}}{5} & \frac{\rho_{VF}}{5} & I_{V}\\
\end{array}
\right),
\end{equation*}
where $I_{K}=1-\sum_{K\neq L} \frac{\rho_{KL}}{5}$, and $K, L\in\{X, Y, E, W, F, V\}$. In the strong imitation limit, the transition matrix is
\begin{equation*}
\left(
\begin{array}{cccccc}
\frac{4}{5}-\frac{2}{5M} & \frac{1}{5} & \frac{1}{5M} & \frac{1}{5M} & 0 & 0\\
0 & \frac{3}{5} & \frac{1}{5} & 0 & \frac{1}{5} & 0\\
\frac{1}{5M} & 0 & 1-\frac{2}{5M} & \frac{1}{5M} & 0 & 0\\
\frac{1}{5M} & 0 & \frac{1}{5M} & 1-\frac{2}{5M} & 0 & 0\\
\frac{1}{5} & 0 & \frac{1}{5} & \frac{1}{5} & \frac{2}{5}-\frac{1}{5M} & \frac{1}{5M}\\
\frac{1}{5} & 0 & \frac{1}{5} & \frac{1}{5} & \frac{1}{5M} & \frac{2}{5}-\frac{1}{5M}\\
\end{array}
\right).
\end{equation*}
Here, the stationary distribution is $[\frac{6}{5M+22},
\frac{3}{5M+22}, \frac{3M+6}{5M+22}, \frac{2M+6}{5M+22},
\frac{3M+1}{(5M+22)(3M+2)}, \frac{1}{(5M+22)(3M+2)}]$, which shows
that peer excluders has an evolutionary advantage over other
strategy individuals (see Fig. S4).

If we consider the second-order exclusion and punishment, then the average payoff for cooperators is given by
\begin{eqnarray*}
\Pi_{X}&=&\frac{\binom{M-E-F-1}{N-1}}{\binom{M-1}{N-1}}[\frac{(N-1)(W\gamma+VB)}{M-1)}+rc-\frac{rc(N-1)Y}{(M-E-F-1)N}-\frac{(N-1)(VB+W\gamma)}{M-E-F-1}]\\
&&-c-\frac{(N-1)(VB+W\gamma)}{M-1}.
\end{eqnarray*}
The payoff for peer excluders is
\begin{eqnarray*}
\Pi_{E}&=&[1-\frac{\binom{M-F-1}{N-1}}{\binom{M-1}{N-1}}][-c-\frac{(N-1)(X+Y+W+V)c_{E}+(N-1)(W\gamma+VB)}{M-1}]+\frac{\binom{M-F-1}{N-1}}{\binom{M-1}{N-1}}\\
&&\sum_{i=0}^{N-1}\sum_{j=0}^{N-1-i}\sum_{p=0}^{N-1-i-j}\sum_{t=0}^{N-1-i-j-p}\frac{\binom{M-F-X-W-Y-V-1}{N-i-j-p-t-1}\binom{X}{i}\binom{Y}{j}\binom{W}{t}\binom{V}{p}}{\binom{M-F-1}{N-1}}[\frac{rc(N-j)}{N-i-j-p-t}-\\
&&t\gamma-pB-(i+j+t+p)c_{E}-c].\nonumber
\end{eqnarray*}
The payoff for peer punishers is
\begin{eqnarray*}
\Pi_{W}&=&-c-\frac{(N-1)(X+Y+E+F)\beta+(N-1)VB}{M-1}-\\
&&\frac{\binom{M-E-F-1}{N-1}}{\binom{M-1}{N-1}}\{-\frac{(N-1)(X+Y+E+F)\beta+(N-1)VB}{M-1}-rc+\\
&&\frac{rc}{N}\frac{(N-1)Y}{M-E-F-1}+\frac{(N-1)[(X+Y)\beta+VB]}{M-E-F-1}\}.
\end{eqnarray*}
The payoff for pool excluders is
\begin{eqnarray*}
\Pi_{F}&=&\sum_{i=0}^{N-1}\sum_{j=0}^{N-i-1}\sum_{p=0}^{N-i-j-1}\sum_{t=0}^{N-i-j-p-1}\frac{\binom{M-E-X-Y-W-V-1}{N-i-j-p-t-1}\binom{X+E}{i}\binom{Y}{j}\binom{V}{p}\binom{W}{t}}{\binom{M-1}{N-1}}\\
&&\times[\frac{r(N-j)c}{N-i-j-p-t}-c-\delta-t\gamma-pB].
\end{eqnarray*}
Last, the payoff for pool punishers is
\begin{eqnarray*}
\Pi_{V}=-c-G-\frac{\binom{M-E-F-1}{N-1}}{\binom{M-1}{N-1}}[\frac{rc}{N}\frac{(N-1)Y}{M-E-F-1}-rc].
\end{eqnarray*}

In the strong imitation limit the transition matrix is given by
\begin{equation*}
\left(
\begin{array}{cccccc}
\frac{2}{5} & \frac{1}{5} & \frac{1}{5} & 0 & \frac{1}{5} & 0\\
0 & \frac{3}{5} & \frac{1}{5} & 0 & \frac{1}{5} & 0\\
0 & 0 & \frac{4}{5} & 0 & \frac{1}{5} & 0\\
0 & 0 & \frac{1}{5} & \frac{3}{5} & \frac{1}{5} & 0\\
0 & 0 & 0 & 0 & 1 & 0\\
0 & 0 & \frac{1}{5} & 0 & \frac{1}{5} & \frac{3}{5}\\
\end{array}
\right).
\end{equation*}
The resulting stationary distribution is $[0, 0, 0, 0, 1, 0]$, which
suggests that pool excluders win (see Fig.~S4).

\section*{9 Robustness of main findings}
\noindent

In the last section we present some representative examples to
illustrate that our main findings are robust and remain valid in a
broad range of model parameters. First, in Fig.~S5 we show the
result of competition between peer exclusion and peer punishment
strategies in the optional PGG for a significantly larger
$\beta=0.7$ fine value. Note that here the fine is almost two times
higher than the punishment cost. Still, excluder strategy remains
dominant in the absence of second-order sanctioning for almost all
imitation strength values. If the second-order punishment and
exclusion are possible then peer exclusion prevails again, as shown
in the right panel of Fig.~S5.

In agreement with our previous findings the superiority of peer
exclusion remains intact no matter how the fine is increased
relevantly in the presence of pool strategies. A typical outcome is
shown in Fig.~S6 where in most of the time the majority of
individuals prefer peer exclusion even if pool strategies are
possible in the absence of second order sanctioning. If the latter
is possible then pool exclusion prevails, as it is demonstrated in
the right panel of Fig.~S6.

Our last figure illustrates the frequencies of available strategies
by choosing different group sizes for the public goods game. Both
panels of Fig.~S7 highlight that group size has no significant role
in the competition of strategies no matter whether second-order
sanctioning is considered or not.

\begin{figure}
\begin{center}
\includegraphics[width=14cm]{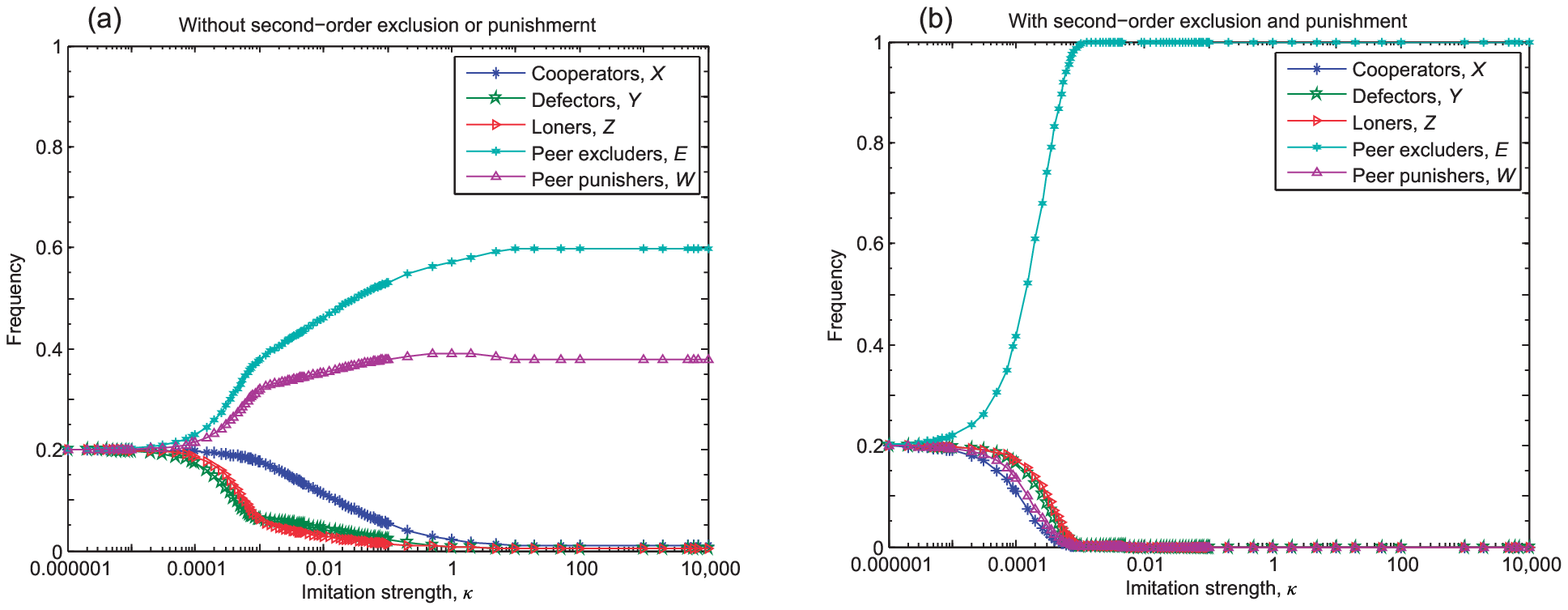}
\caption{(Color online) The competition between peer exclusion and
peer punishment in the optional PGG for a larger punishment fine
$\beta=0.7$. In the absence of second-order sanctioning, shown in
panel~(a), both strategies survive but peer exclusion dominates. If
second-order exclusion and punishment are applied then peer
excluders prevail, as shown in panel~(b). Parameters: $N=5$, $r=3$,
$c=1$, $\mu=10^{-6}$, $\sigma=1$, $M=100$, $c_{E}=0.4$, and
$\gamma=0.4$.} \label{figs5}
\end{center}
\end{figure}

\begin{figure}
\begin{center}
\includegraphics[width=14cm]{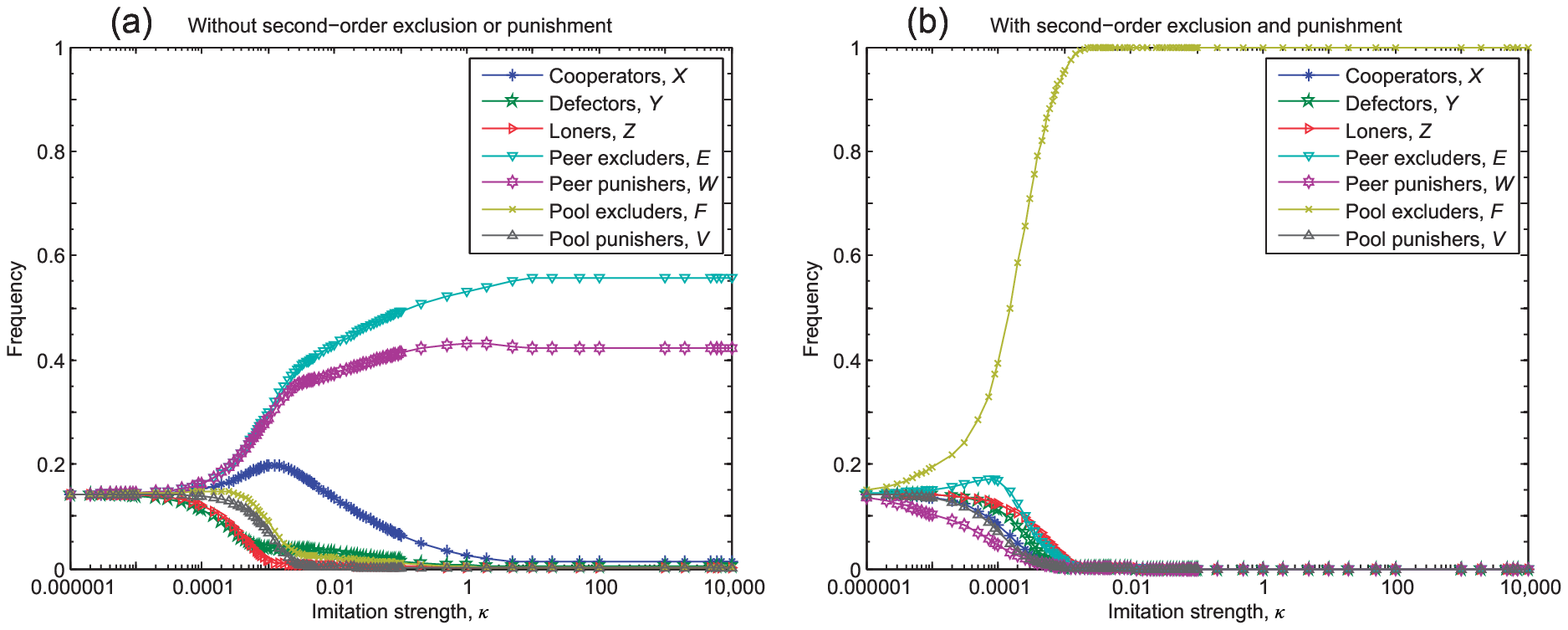}
\caption{(Color online) The competition between different types of
social exclusion and costly punishment in the optional PGG for a
large punishment fine $\beta=0.7$. When strategies can penalize
defectors only, shown in panel~(a), then all strategies coexist in
time average for weak strength of imitation, but in most of the time
peer excluders form the majority of the population for other
imitation strength values. Panel~(b) shows the case when
second-order sanctioning is present. Here pool excluders prevail and
dominate the whole population. Parameters: $N=5$, $r=3$, $c=1$,
$\mu=10^{-6}$, $\sigma=1$, $M=100$, $c_{E}=\delta=0.4$,
$\gamma=0.4$, and $B=G=0.4$.} \label{figs6}
\end{center}
\end{figure}

\begin{figure}
\begin{center}
\includegraphics[width=14cm]{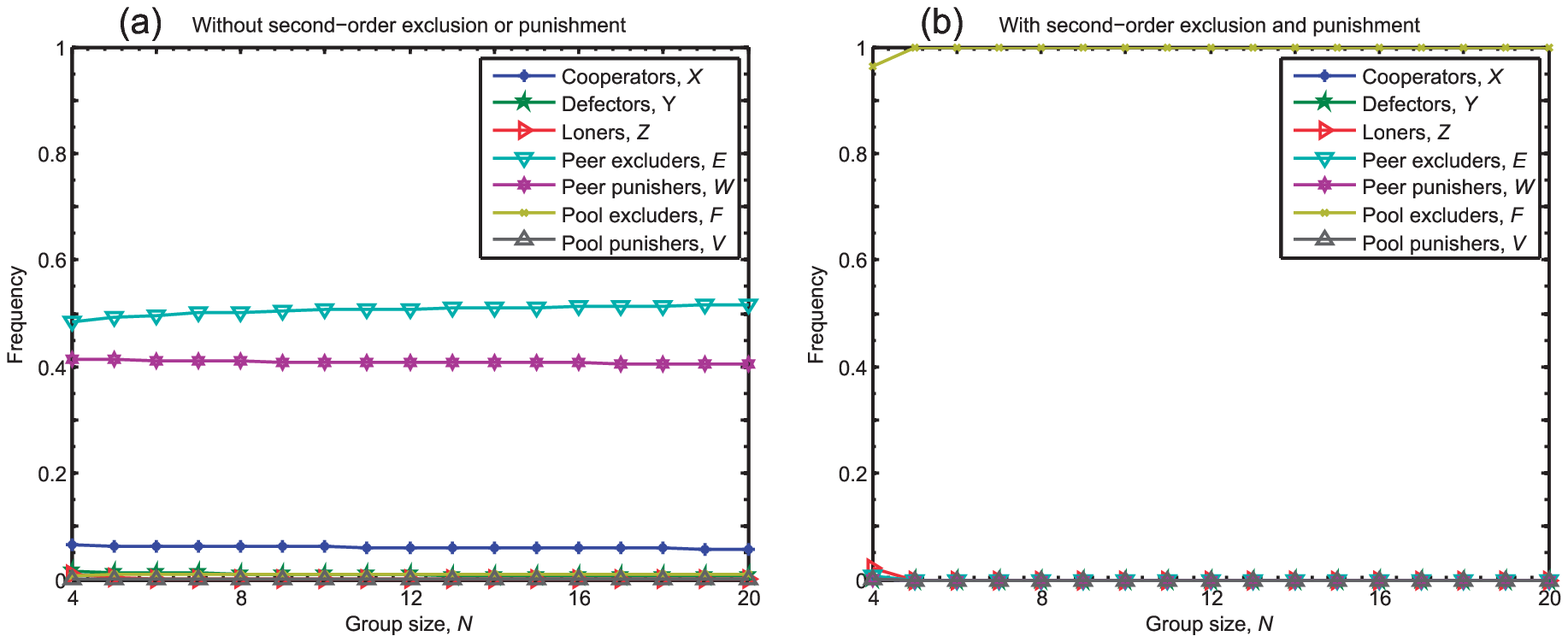}
\caption{(Color online) The competition between different types of
social exclusion and costly punishment in the optional PGG as a
function of the group size $N$ for imitation strength $\kappa=0.1$.
It is shown that group size has no significant role in the
competition of strategies no matter whether second-order sanctioning
is considered or not. Parameters: $r/N=0.6$, $c=1$, $\mu=10^{-6}$,
$\sigma=1$, $M=100$, $c_{E}=\delta=0.4$, $\beta=\gamma=0.4$, and
$B=G=0.4$.} \label{figs7}
\end{center}
\end{figure}